# An all-weather microwave-based AI retrieval of pan-Arctic melt pond fraction

Sara Aparício[a], Simon Driscoll[b], Mark R. Drinkwater[c], Daniela Flocco[d,e]

[a]CENSE - Center for Environmental and Sustainability Research & CHANGE - Global Change and Sustainability Institute, NOVA School of Science and Technology, Universidade Nova de Lisboa, Campus de Caparica, 2829-516 Caparica, Portugal
[b]Department of Applied Mathematics and Theoretical Physics, University of Cambridge, United Kingdom
[c]ViaEO Consulting, Wassenaar, The Netherlands
[d]Dipartimento di Scienze della Terra, dell'Ambiente e delle Risorse, Università degli Studi di Napoli "Federico II", Napoli, Italy
[e]CoNISMA - Consorzio Nazionale Interuniversitario per le Scienze del Mare, Roma, Italy

*Correspondence to:* Sara Aparício (s.aparicio@campus.fct.unl.pt)

**Abstract**:

Melt pond fraction (MPF) is a key variable governing Arctic sea ice albedo and seasonal sea ice predictability, yet existing pan-Arctic MPF products rely almost exclusively on optical sensors, limiting their availability under cloud cover and seasonal darkness. This study presents PRISM-MPF, the first pan-Arctic, multi-year, all-weather MPF dataset, derived from PRISM (Polar Retrievals through Intelligent Sensor Merging), a machine learning framework fusing multi-source microwave observations with auxiliary thermal and spatiotemporal information. PRISM-MPF is designed to complement optical products and support Arctic monitoring under all atmospheric conditions. A gradient-boosted decision-tree model (XGBoost) was trained on eight melt seasons (2017-2025, excluding 2019) using C-band SAR backscatter from Sentinel-1, passive microwave brightness temperatures and sea ice concentration from AMSR2, sea surface temperature from NOAA's Optimum Interpolation SST (OISST), and spatiotemporal descriptors (date and grid position). Three experiments assessed baseline skill and feature importance, independent sensor-group contributions (including CryoSat-2 radar freeboard), and regional specialisation, all under Leave-One-Year-Out (LOYO) cross-validation on unseen melt seasons. The baseline PRISM retrieval achieved a per-season mean $R^2$=0.468 and RMSE=0.080 (pooled over all 1,066,808 pixels: $R^2$=0.532, RMSE=0.077), comparable to prior regional microwave-based studies, now at pan-Arctic, multi-year scale. Removing spatiotemporal features or C-band SAR reduced skill comparably ($\Delta R^2$=+0.089 and +0.085), whereas physical sensors alone outperformed spatiotemporal features alone ($R^2$=0.375 vs 0.261), showing that the two groups are complementary. Regional specialisation offered no advantage over a single pan-Arctic retrieval. This framework establishes a reproducible foundation for all-weather Arctic MPF monitoring, extensible to new sensors, longer records, and reference products.

## 1. Introduction

Melt ponds are a major contributor to the ice-albedo feedback that drives the rapid transformation of the Arctic summer surface (Fig. 1). Substantially darker than the surrounding snow and bare ice, they enhance solar absorption and accelerate melt, playing a central role in the summer surface energy budget (Polashenski et al., 2012; Pistone et al., 2014). For this reason, melt pond fraction (MPF), the fraction of sea ice area covered by ponds, is a key variable for parameterising sea ice albedo in climate and sea ice models (Flocco and Feltham, 2007; Hunke et al., 2013), and late-spring MPF carries predictive skill for September sea ice extent (Schröder et al., 2014; Liu et al., 2015). Robust, spatially and temporally comprehensive MPF observations are thus essential for both process understanding and seasonal sea ice forecasting.

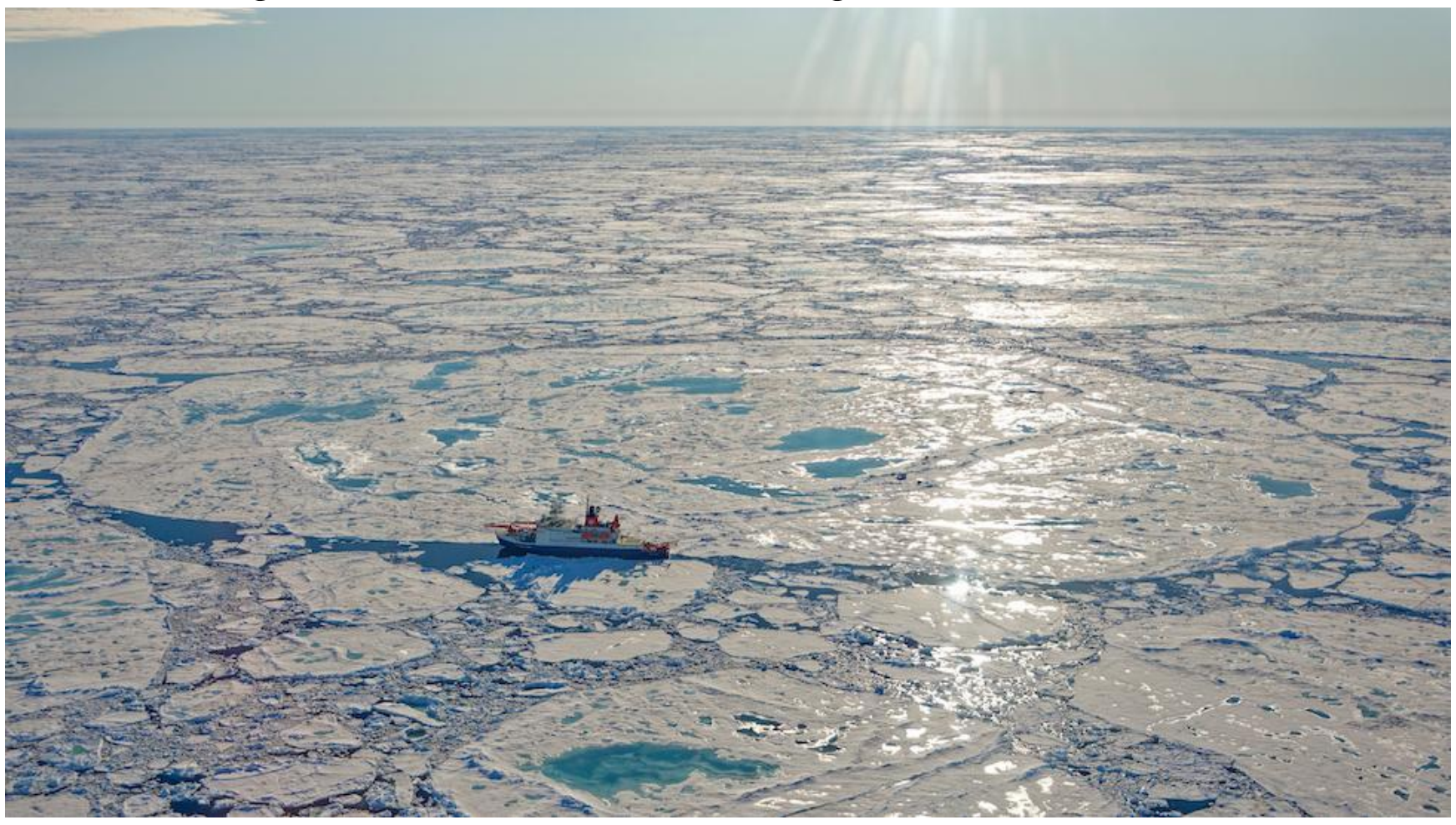

**Figure 1:** Arctic melt ponds, with the German icebreaker RV Polarstern for scale. Photo credit: MOSAiC/Alfred Wegener Institute

A systematic review, which analysed over forty Earth observation (EO) studies of Arctic melt ponds (Aparício et al., 2025), underscored a noticeable asymmetry in the field of EO of Arctic melt ponds. Current multi-year pan-Arctic MPF products are predominantly based on optical sensors such as the Moderate Resolution Imaging Spectroradiometer (MODIS), the Medium Resolution Imaging Spectrometer (MERIS), and the Sentinel-3 Ocean and Land Colour Instrument (OLCI) (Ding et al., 2020; Lee et al., 2020; Peng et al., 2022). Microwave sensors have not achieved comparable operational reliability for MPF retrieval, despite their all-weather, day-and-night observational advantage. Synthetic Aperture Radar (SAR), passive microwave radiometry, and radar altimetry have each been applied to melt pond studies: for pond detection from backscatter contrast (Yackel & Barber, 2000; Scharien et al., 2014; Scharien et al., 2017; Han et al., 2016), basin-scale MPF estimation from brightness temperature spatial and spectral gradients (Tanaka et al., 2016; Tanaka & Scharien, 2022), and characterisation of pond-induced freeboard biases and melt detection (Kwok et al., 2018; Landy et al., 2022), respectively. These studies were conducted over specific regions, primarily landfast first-year ice (FYI) in the Canadian Arctic Archipelago (CAA), Beaufort and Chukchi Seas, and focused on one or a small number of melt seasons (Aparício et al., 2025). Beyond these regional and temporal constraints, microwave-based MPF retrieval also faces physical limitations: discriminating between the SAR backscatter signature of melt ponds and that of open water is challenging and highly sensitive to wind conditions and incidence-angle (Yackel & Barber, 2000; Li et al., 2017). Passive microwave brightness temperature ($T_B$) also suffers from ambiguity between melt pond signatures and open ocean (Tanaka et al., 2016). Radar altimetry has historically treated melt pond specular returns as a source of freeboard bias rather than an exploitable signal (Landy et al., 2022). As a result, no pan-Arctic, multi-year, cross-validated microwave-based MPF product currently exists at a scale comparable to established optical products.

This study answers to the interest of a multi-sensor fusion strategy that combines multiple microwave observations followed by the application of a machine learning algorithm to retrieve MPF, an approach that has been largely applied in the optical domain but less so in the microwave domain, where most of existing work relies on regression against a single or dual sensor combination (e.g., Tanaka et al., 2016; Ramjan et al., 2018). Addressing this gap, we present PRISM-MPF, the first pan-Arctic, multi-year all-weather MPF dataset, derived from PRISM (Polar Retrievals through Intelligent Sensor Merging), an XGBoost-based machine learning framework for MPF retrieval. PRISM fuses four independent, complementary data sources: C-band SAR backscatter (Sentinel-1), passive microwave brightness temperature and sea ice concentration (AMSR2), sea surface temperature (OISST), and radar freeboard (CryoSat-2), which are trained and evaluated across eight melt seasons (2017-2025, excluding 2019) using a conservative cross-validation protocol designed to test generalisation to entirely unseen melt seasons.

The remainder of this paper is organised as follows. Section 2 describes the study area. Section 3 presents the datasets used as input and reference variables. Section 4 details the data preprocessing, spatial harmonisation, and feature engineering steps used to construct the training datasets, describes the model architecture and cross-validation design, and outlines the three experimental configurations: a baseline pan-Arctic retrieval algorithm evaluation (Experiment A), a systematic sensor and feature contribution analysis (Experiment B), and a spatial specialisation assessment (Experiment C). Results are presented in Section 5, followed by discussion of their implications for microwave-based MPF retrieval and directions for future work in Section 6.

## 2. Study Area

The study area encompasses the entire Arctic sea-ice domain and was divided into geographical regions to assess retrieval performance across individual regions (Experiment A2; Section 4.5) and to determine whether region-specific retrievals outperform a single pan-Arctic retrieval algorithm (Experiment C; Section 4.5). The nine sectors defined are: the Central Arctic, the Canadian Arctic Archipelago (CAA), the Beaufort Sea, the Chukchi Sea, the East Siberian Sea, Laptev Sea, Kara Sea, Barents Sea and the Greenland Sea (Fig. 2). This partitioning follows the conventional division of the Arctic Ocean into its major marginal seas and broadly corresponds to the subregions adopted in previous Arctic-wide studies across different physical domains, including sea-ice albedo (Peng et al., 2019), surface wind and wave climate (Cabral et al., 2022), and ocean biogeochemical dynamics (Manizza et al., 2023). Beyond using similar geographical conventions, these studies demonstrate that the regions exhibit distinct physical characteristics and dynamics. Moreover, melt pond characteristics vary with ice type: ponds tend to be shallow and extensive on smooth FYI, and smaller and deeper on topographically constrained multiyear ice (MYI) (Polashenski et al., 2012). As the proportions of these ice types differ markedly between sectors, the subdivision also reflects distinct melt pond regimes rather than geographical convenience alone, and is complemented by the latitude-band and ice concentration strata defined in Experiments A2 and C (Section 4.5).

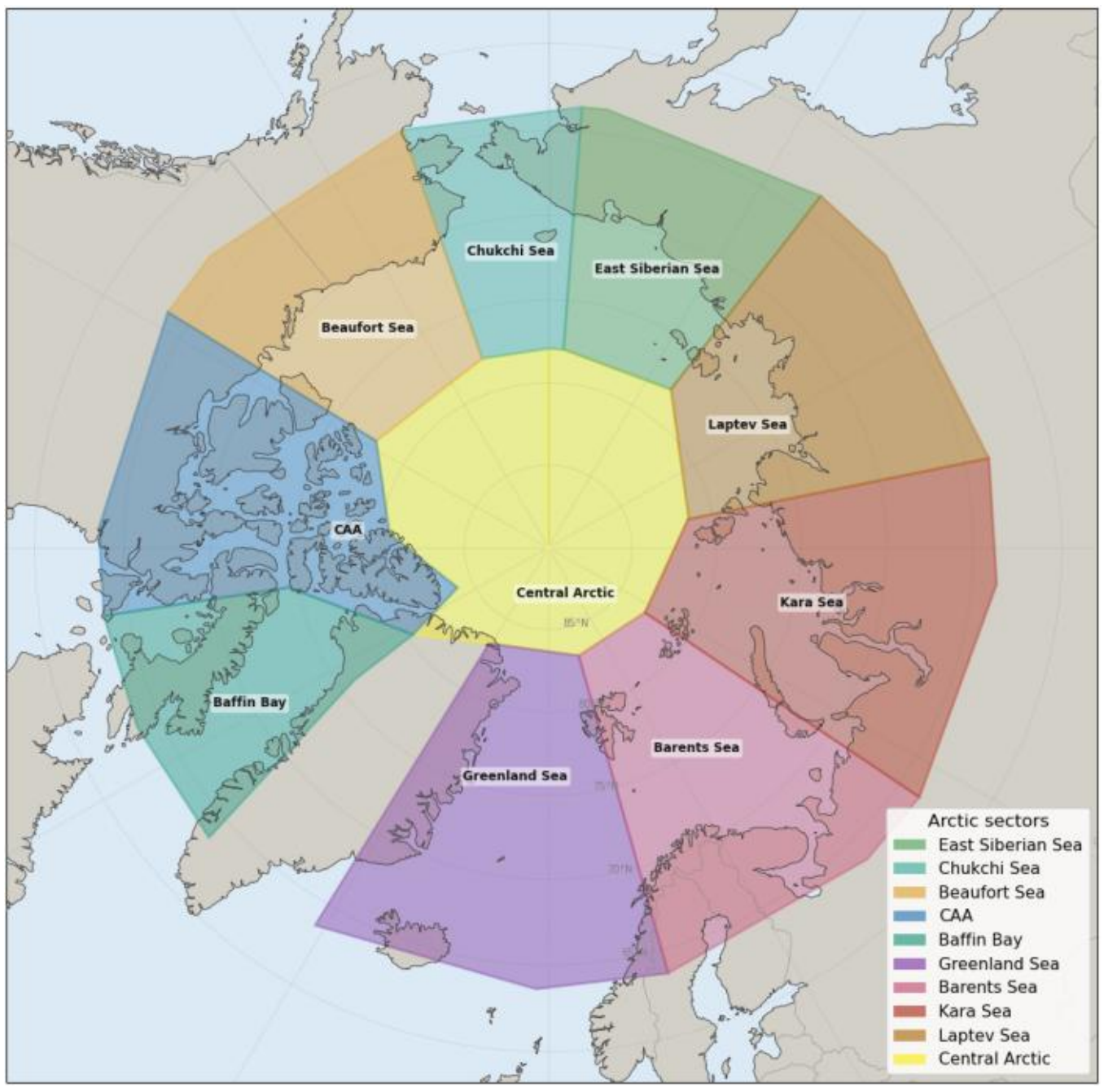

**Figure 2:** Arctic geographic sectors

## 3. Data

The aim of this study is to predict MPF based on multiple independent input variables. The target variable is the OLCI-Istomina MPF product (Istomina et al., 2025), an optically derived, pan-Arctic MPF dataset described in full in Section 3.5, which serves as the training target from which the ensemble machine learning XGBoost algorithm models the microwave-MPF relationship. The input variables are drawn from two sources: satellite-derived observations and spatiotemporal descriptors. The satellite inputs comprise C-band SAR backscatter from Sentinel-1; passive microwave brightness temperature and ice concentration from AMSR2; sea surface temperature from OISST, and radar freeboard from CryoSat-2; and a set of spatiotemporal variables derived from each pixel's grid position and acquisition date, encoding seasonal and geographic context. CryoSat-2 radar freeboard was incorporated as an additional input source on a matched pixel subset, given its sparse spatial coverage relative to the other datasets. Note that these are the raw data sources from which the actual model predictors, such as polarisation ratios, cyclical date encodings, and other derived variables, were engineered, as described in Section 4.2. Each dataset source is described in more detail below.

### 3.1. SAR imagery (from Sentinel-1)

The Copernicus Sentinel-1 (S1) first generation mission comprises a constellation of three sun-synchronous polar-orbiting satellites (S1-A, S1-C and S1-D), launched respectively in 2014, 2024 and 2025. The second unit (S1-B), launched in 2016, failed in orbit and stopped delivering data in December 2021. The S1 mission carries a C-band SAR, which enables acquisition of imagery regardless of the weather or daylight conditions. Its SAR instrument operates in four imaging modes and provides dual-polarised radar backscatter. For this study, backscatter ($\sigma^0$) from the HH co-polarised and HV cross-polarised channels were used for the period 2017-2025, with the exclusion of 2019, due to insufficient HV S1 coverage during that year. In Extra Wide (EW) swath mode (the one used in this study), S1 provides a spatial resolution of about 90 m and revisits most of the Arctic every 1-3 days, allowing it to resolve the surface roughness and water-ice contrast associated with melt ponds, which range from metres to tens of metres in size and evolve over days to weeks (Polashenski et al., 2012). Data were accessed through Google Earth Engine (GEE) (COPERNICUS/S1_GRD collection), which provides S1 Level-1 Ground Range Detected (GRD) scenes, for which several preprocessing steps had already been applied, namely orbit file application, GRD border noise removal, thermal noise removal, radiometric calibration and terrain correction.

### 3.2. Brightness Temperature & Sea Ice Concentration (from AMSR2)

The Advanced Microwave Scanning Radiometer 2 (AMSR2) onboard JAXA's Global Change Observation Mission for Water (GCOM-W1), was launched in 2012 and acquires passive microwave brightness temperature ($T_B$) at multiple frequencies between 6.925 and 89.0 GHz with vertical and horizontal polarisations. Although its footprints are coarse (from about 22 × 14 km at 18.7 GHz to about 5 × 3 km at 89 GHz), AMSR2 provides daily pan-Arctic coverage, and its brightness temperatures are sensitive to the liquid water and wet snow within each footprint (Kern et al., 2016), complementing the finer but less frequent SAR observations. The data accessed via NASA Earthdata comprised the AMSR-E/AMSR2 Unified L3 Daily 12.5 km Brightness Temperatures and Sea Ice Concentration product (AU_SI12). AU_SI12 combines the AMSR2 observations with those of the predecessor mission AMSR-E, onto a consistent 12.5 x 12.5 km² polar stereographic grid. This dataset also contains sea ice concentrations, retrieved with the Enhanced NASA Team (NT2) algorithm (Meier et al., 2018).

### 3.3. Sea Surface Temperature (from OISST)

The NOAA Optimum Interpolation Sea Surface Temperature (OISST) version 2.1 is a long-term Climate Data Record that incorporates blended observations from different platforms (satellites, ships, buoys and Argo floats) into a regular global grid (Huang et al., 2021). The dataset is interpolated to fill gaps, creating a spatially complete 0.25° × 0.25° gridded map of SST. Satellite and ship observations are referenced to buoys to compensate for platform differences and sensor biases. Since satellites do not observe under sea ice, OISST v2.1 assigns proxy SST values in ice-covered cells.  derived from the seven-day median sea ice concentration and set close to the climatological seawater freezing temperature under consolidated ice (Banzon et al., 2020). Thus, within the pack, OISST SST largely reflects ice concentration rather than an independent temperature measurement, carrying genuine thermal information mainly near the ice edge and in open water. OISST data was selected to provide thermal context on the ocean and ice-edge conditions of ponded ice. The dataset was accessed via GEE (NOAA/CDR/OISST/V2_1 collection).

### 3.4. Radar freeboard (from CryoSat-2)

The CryoSat-2 (CS) mission, launched in 2010, carries a SAR/Interferometric Radar Altimeter (SIRAL) on board, operating in the Ku-band at 13.575 GHz (Drinkwater et al., 2004; Wingham et al., 2006). In SAR mode, CryoSat-2 has a nominal synthesized footprint of approximately 0.31 × 1.67 km in the along- and across-track directions, respectively (Kwok and Morison, 2016), providing detailed but spatially sparse sampling. Its radar returns can be affected by specular reflections from melt pond surfaces. Arctic radar freeboard data, i.e. the height of the snow-ice interface above the local sea level as measured by radar ranging, were obtained from the Cryo-TEMPO Summer Sea Ice thematic product (TEMPO_POCA_SS; ESA, 2024; Braakmann-Folgmann, 2024), accessed through the ESA CryoSat-2 Science Server (science-pds.cryosat.esa.int) in July 2025. The product is now distributed through the ESA CS2EO platform (https://cs2eo.org/) and catalogued on ESA Earth Online (https://earth.esa.int/eogateway/catalog/cryotempo). Files were obtained as along-track Level-2 products in NetCDF format, each containing per-point latitude, longitude, radar freeboard, smoothed radar freeboard, and retrieval uncertainty.

### 3.5. Reference melt pond fraction (from OLCI-Istomina)

The OLCI-Istomina MPF product serves as the reference for this study: it provides the training target for PRISM and the benchmark against which PRISM-MPF is evaluated, while carrying its own retrieval uncertainty (Section 6.2). This product is a satellite-derived 300 m gridded dataset that provides daily MPF estimates over Arctic sea ice using data from ENVISAT MERIS and Sentinel-3 OLCI. The product builds on a spectral unmixing algorithm originally developed for MERIS data (Istomina et al., 2015a, 2015b) and extended to the full 2002-2025 record incorporating Sentinel-3 OLCI (Istomina et al., 2025). It applies spectral unmixing approaches for melt-pond detection in clear-sky, daytime multichannel optical image data by exploiting blue/green/NIR spectral contrasts to separate melt ponds from bare ice. OLCI-Istomina was selected as the PRISM-MPF training target for three reasons. First, it is a physically based, pan-Arctic MPF dataset designed for internally consistent large-scale applications, a prerequisite for training a pan-Arctic model; second, compared to MODIS-based products, it offers a stronger spectral constraint and avoids the saturation limitations reported for bright Arctic sea ice and melt pond surfaces, whilst providing better methodological consistency with the optical complexity of wet snow, ponded ice, and open water during the melt season; and third, the product has been validated against Sentinel-2 MSI imagery (Istomina et al., 2025), providing independent evidence of its retrieval quality and making it a robust training target for the PRISM framework. As an optically derived product requiring clear-sky, daytime conditions, OLCI-Istomina is complementary to rather than competing with PRISM-MPF: it provides the high-quality training labels from which the all-weather microwave retrieval learns, whilst PRISM-MPF aims to extend MPF monitoring to the cloud-covered periods that optical sensors cannot observe.

## 4. Methods

### 4.1. Data preprocessing and spatial harmonisation

Each dataset underwent specific preprocessing and feature engineering steps, before proceeding to the final co-location step (Figure 3). The MPF product uses the EPSG:3413 polar stereographic National Snow and Ice Data Center (NSIDC) 12.5 x 12.5 km grid and is distributed in NetCDF format, requiring format conversion to GeoTIFF only, making it the master reference to which all other datasets were subsequently aligned. MPF values are stored as integers from 0 to 100, representing the percentage of each grid cell covered by melt ponds, and were scaled to fractional units in the range [0, 1]. Since the NetCDF files carried no embedded coordinate reference system (CRS), the EPSG:3413 polar stereographic CRS and the NSIDC affine transform, defined by an upper-left corner at (−3,843,750 m, +5,843,750 m) and a 12,500 m pixel spacing over a 896 x 608 grid, were assigned at write time. Before writing to the output GeoTIFF, the array was also flipped to correct row orientation, since the source file stores rows from South to North, whereas the affine transform assigns row '0' to the northernmost coordinate.

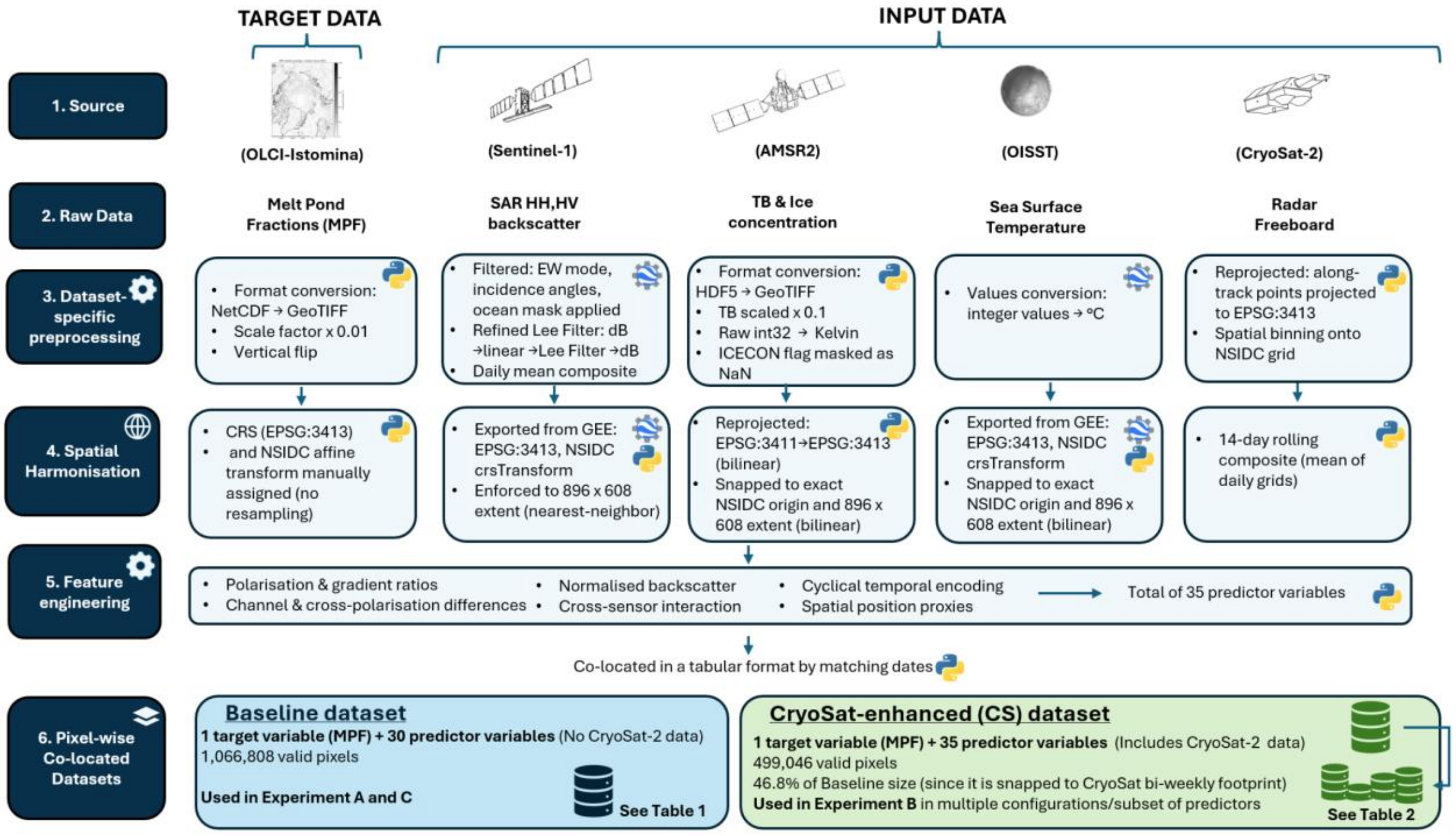

**Figure 3:** Dataset preparation and pixel-wise colocation workflow

Sentinel-1 data acquisitions were filtered for Extra Wide Swath (EW) mode, HH and HV polarisations, and both ascending and descending passes, with incidence angles between 31° and 45°, on GEE. In addition to the Level-1 GRD processing applied by GEE, a Refined Lee filter (Lee et al., 1999) was applied to every image, beginning with the conversion of radar backscatter values from decibels (dB) to natural units (power domain) using the transformation:

$$\sigma^0 = 10^{\left(\frac{dB}{10}\right)} \quad \textbf{(1)}$$

After application of the Refined Lee filter, the backscatter values were transformed back to decibels using the inverse transformation:

$$dB = 10 \cdot log_{10}(\sigma^0) \quad \textbf{(2)}$$

A daily mean composite was computed from all available S1 acquisitions for the same day, resulting in a continuous coverage of overlapping areas in a single band. Each daily composite was exported from GEE in EPSG:3413 using the NSIDC affine transform, ensuring the same grid origin as the master reference grid. Locally, the exports were resampled to the exact 896 × 608 NSIDC extent using nearest-neighbour resampling.

OISST data, also accessed via GEE, were filtered to one daily image per date, rescaled from stored integer values to physical units by dividing by 100 to obtain sea surface temperature in °C, and masked to ocean-only pixels using the same land mask applied to S1. The dataset's native resolution of 0.25° was resampled to a 12.5 x12.5 km grid spacing during GEE export, using the same NSIDC coordinate transform as the master reference grid. Since the GEE exports were clipped to a specific circular Area Of Interest (AOI), the resulting files had a different origin and extent from the NSIDC grid. A local grid-snapping step was therefore applied (bilinear resampling) to enforce the exact grid and origin, completing alignment with the master reference.

AMSR2 brightness temperature channels (18.7, 36.5 and 89.0 GHz) in both H and V polarisations were extracted from the HDF5 files and rescaled (×0.1) converting raw integer values to Kelvins, while the sea ice concentration band was read directly as percentage (0-100) without scaling, with flag values representing land (120) and missing data (110) set to not-a-number (NaN). As AMSR2 is distributed on the EPSG:3411 polar stereographic grid (Hughes 1980 ellipsoid) with an origin at (−3,850,000 m, +5,850,000 m), it was firstly reprojected from EPSG:3411 to EPSG:3413 using bilinear resampling, with NaN explicitly set as the source and destination no-data value to prevent interpolation from spreading masked values into valid neighbouring cells; and subsequently snapped to the exact 896 × 608 NSIDC extent and origin, also using bilinear resampling.

Finally, since CryoSat-2 data consisted of daily along-track NetCDF files, they were processed in two steps. Three variables (radar freeboard, smoothed radar freeboard, and retrieval uncertainty), were first projected from geographic coordinates (lat/lon) to EPSG:3413, after which they were spatially binned onto the master NSIDC 12.5 km grid using a mean statistic per cell. Grid cells with no along-track passes on a given day were written as no-data. In a second step, given that CryoSat-2's daily coverage of the Arctic is only approximately 15% due to its narrow altimeter swath, daily grids were aggregated into 14-day rolling composites by computing the pixel-wise mean across all daily grids within the date and the two prior weeks, to enhance spatial coverage. Grid cells with no valid passes within the 14-day window remained as NaN and were excluded.

### 4.2. Building training data with feature engineering and co-location

In a final step, tabular training data were created through pixel-wise colocation of the harmonised data products described above. Prior to this, feature engineering was performed to harness extra information from band combinations, grid geometry and acquisition dates (see Table 1). From the Sentinel-1 SAR co-polarised (HH) and cross-polarised (HV) backscatter coefficient data, two additional predictors were derived: normalised HV backscatter and the cross-polarisation difference:

$$Normalised\ HV = \frac{\sigma^0(dB)\ HV}{\sigma^0(dB)\ HH} \quad \textbf{(3)}$$

$$\textit{Cross-polarisation difference} = \sigma^0(dB)\ HH - \sigma^0(dB)\ HV \quad \textbf{(4)}$$

Since $\sigma^0$ is expressed in dB, the cross-polarisation difference corresponds to the HH/HV ratio in linear units, whereas the normalised HV is an empirical ratio of logarithmic values without direct physical interpretation. From the AMSR2 brightness temperatures, nine spectral features were computed. Polarisation ratios at 18, 36 and 89 GHz:

$$\boldsymbol{PR(f)} = \frac{\boldsymbol{T_b(f,V) - T_b(f,H)}}{\boldsymbol{T_b(f,V) + T_b(f,H)}}, \boldsymbol{f} \in \{\boldsymbol{18, 36, 89}\}\ \boldsymbol{GHz} \quad \textbf{(5)}$$

Gradient ratios between frequency pairs:

$$\boldsymbol{GR}(f_1, f_2, V) = \frac{\boldsymbol{T_b}(f_1,V) - \boldsymbol{T_b}(f_2,V)}{\boldsymbol{T_b}(f_1,V) + \boldsymbol{T_b}(f_2,V)}, (f_1, f_2) \in \{(36,18), (89,36), (89,18)\}\ \text{GHz} \quad \textbf{(6)}$$

Channel difference terms:

$$\Delta T_b(f_1, p_1, f_2, p_2) = T_b(f_1, p_1) - T_b(f_2, p_2), \quad (f_1, p_1, f_2, p_2), \in \{(89V, 36V), (89H, 36H), (36V, 18V)\} \quad \textbf{(7)}$$

Three cross-sensor interaction terms were also computed by multiplying SAR HH backscatter, SAR HV backscatter, and the 18 GHz polarisation ratio by the ice concentration fraction $c$:

$$\boldsymbol{I_k} = \boldsymbol{xk} \cdot \boldsymbol{c}, \quad \boldsymbol{k} \in \{\boldsymbol{\sigma^0}HH, \sigma^0HV, PR(18)\} \quad \textbf{(8)}$$

where $x_k$ is the sensor feature and $c = icecon_frac \in [0,1]$ is the sea ice concentration fraction.

Finally, five spatiotemporal features were derived from each pixel's grid position and acquisition date rather than from any sensor: day-of-year encoded cyclically as sine and cosine components (doy_sin, doy_cos) to avoid a discontinuity between 31 December and 1 January; calendar month (month); and normalised grid row and column (row_norm, col_norm), which serve as continuous proxies for latitude and longitude within the NSIDC grid. The full feature set is listed in Table 1.

Two training datasets, “Baseline” and “CryoSat enhanced”, were subsequently created (see the last panel of Figure 3). The baseline dataset contains the target variable and all input data from S1, AMSR2 and OISST together with the 30 features listed in Table 1. This dataset retains every pixel where all three input sources and the MPF target are simultaneously valid, and totals 1,066,808 pixels spanning the 2017-2025 melt seasons (June-August, excluding 2019). The CryoSat-enhanced training dataset additionally includes 14-day rolling composite CryoSat-2 data. Pixels with no valid CryoSat-2 observations within the composite window were retained as NaN rather than excluded, preserving the spatial extent of the dataset. Pixels requiring simultaneously valid CryoSat-2 observations are substantially reduced given that the 2-week mean coverage is still sparse, reducing this second tabular training dataset to 499,046 pixels (47% of the baseline). This trade-off, i.e., gaining a physically informative predictor at the cost of a substantially smaller training set, is the reason CryoSat-2 was isolated into a separate dataset rather than merged into the baseline. Both datasets were used in different experimental configurations: the baseline dataset in Experiments A and C, and the CryoSat-2 dataset in Experiment B, as described in full in the experimental design (Section 4.5).

### 4.3. Model design (XGBoost)

For our model, we selected Gradient Boosting which works by sequentially adding decision trees to an ensemble, each one correcting the residual error of its predecessor (Friedman, 2001), implemented with the Python library package eXtreme Gradient Boosting (XGBoost; Chen and Guestrin, 2016). XGBoost was selected for its scalability to large tabular datasets. It has the capability to handle nonlinear relationships between heterogeneous variables (as is the case of microwave radar backscatter values, brightness temperature, or sea-surface temperature); and finally, for its built-in L1/L2 regularisation, that prevents correlated features groups (as is the case of feature engineered ratios and gradients) from dominating without need of manual selection. The number of trees (800) and learning rate (0.03) were set to provide a conservative regularisation while minimum child weight (10) was chosen to prevent splits on rare high MPF (i.e. <5% pixels). These and all remaining hyperparameters were fixed for all experiments to allow comparison (Table A in Appendix). This model learns from raw and feature-engineered input, as a supervised regression task, as is the case of predicting continuous values of MPF bounded to [0, 1]. To enforce this physical constraint, XGBoost was configured with a bounded regression objective that constrains all predictions to the interval (0, 1) by applying a sigmoid function to the raw ensemble output:

$$\hat{y} = \frac{1}{1 + e^{-f(x)}} \quad (9)$$

where $f(x)$ is the raw ensemble output and $\hat{y}$ is the predicted PRISM-MPF value, bounded to the open interval $(0,1)$. This ensures all retrievals remain within physically valid MPF bounds without requiring any transformation of the training labels. Once trained, the XGBoost model constitutes the PRISM retrieval, and the MPF values it produces form the PRISM-MPF dataset. Throughout this paper, "model" refers to the XGBoost architecture and its training, "PRISM" to the trained retrieval whose skill and behaviour are evaluated, and "PRISM-MPF" to the retrieved MPF dataset.

### 4.4. Cross validation

A conservative evaluation was implemented to assess the ability of the retrieval algorithm to generalise to entirely unseen melt seasons (i.e., to maintain predictive skill on data not used in training), while avoiding temporal leakage (i.e., the model implicitly benefiting from information about the period being predicted). Retrieval performance was assessed using GroupKFold, with years as the grouping variable, following a Leave-One-Year-Out cross validation (LOYO CV) approach. For each fold, one full melt season (June-August) of a specific year is held out as a validation set while all the remaining years were used to train the model. This ensures that no pixel from a given year influences the training of the model used to predict that same year. Out-of-fold (OOF) predictions were assembled by concatenating the held-out predictions from each fold covering all data pixels, where each of them was made by a model that had never seen that pixel's year. All performance metrics are computed from these OOF predictions. In addition, a final XGBoost model was trained on all eight melt seasons using the same hyperparameters; this model constitutes the PRISM-MPF dataset and for the SHAP feature importance analysis (Section 5.1), while its expected skill on unseen years is given by the LOYO OOF evaluation. Retrieval performance was quantified by complementary metrics:

$$R^2 = 1 - \frac{\sum_i (y_i - \hat{y}_i)^2}{\sum_i (y_i - \bar{y})^2} \quad (10)$$

$$RMSE = \sqrt{\frac{1}{N}\sum_i (y_i - \hat{y}_i)^2} \quad \textbf{(11)}$$

$$Mean\ bias = \frac{1}{N}\sum_i (\hat{y}_i - y_i) \quad \textbf{(12)}$$

Where $\boldsymbol{y_i}$ is the observed MPF from the reference product, $\hat{\boldsymbol{y}}_i$ is the MPF predicted by the retrieval algorithm, $\bar{\mathbf{y}}$ is the mean observed MPF, and *N* is the number of pixels. $R^2$ is used to quantify how much of MPF variance is explained by the retrieval (and specifically the contribution of each sensor and feature) while root mean square error (RMSE) is used to quantify the absolute prediction error and to compare to existing MPF retrieval studies. The mean bias quantifies systematic over- (positive) or underestimation (negative) of MPF, and its square indicates the share of the mean squared error ($RMSE^2$) due to this systematic offset.

### 4.5. Experiment design overview

Three sets of experiments were conducted to address different questions (Fig. 4). Experiment A establishes the baseline PRISM retrieval algorithm (a single pan-Arctic XGBoost model trained on 30 features) and assesses how well a single pan-Arctic retrieval algorithm generalises to unseen melt seasons, that is, provides robust predictions for melt seasons not included in the training dataset, including an examination of how skill varies by region, latitude, and ice concentration regime (Experiment A1: global retrieval evaluation and feature importance; Experiment A2: stratification diagnostics). Experiment B quantifies the independent contribution of each sensor and feature group to retrieval performance, isolating whether the model is primarily learning from physical sensor signals, spatiotemporal patterns, or their combination. Experiment C tests whether the pan-Arctic retrieval algorithm can be improved by spatial specialisation through training dedicated regional models (Experiment C1: local versus global performance; Experiment C2: per-stratum feature importance).

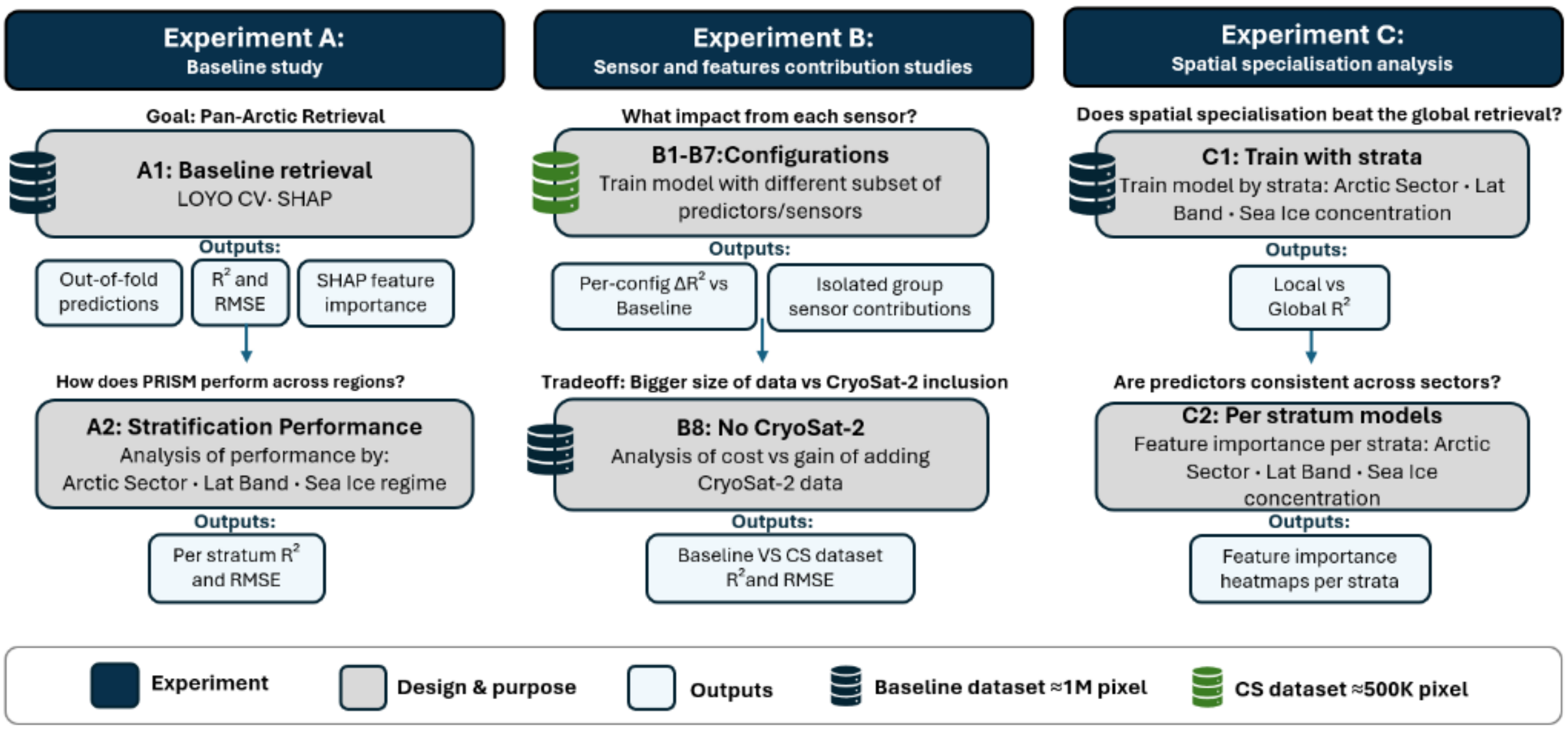

**Figure 4:** Experiments design, purposes and outputs

***Experiment A - Baseline study: pan-Arctic retrieval evaluation***

Experiment A trains the baseline XGBoost model using all input features and evaluates the resulting PRISM retrieval algorithm under LOYO cross-validation: in each fold, one melt season (June-August) is withheld entirely from training and used only for testing. This assesses how well the retrieval algorithm predicts MPF in a year it has never seen, which is the situation it faces when applied to new melt seasons. This experiment uses all 30 base and feature-engineered variables described in Section 4.2 and listed in Table 1, excluding CryoSat-2-derived features. Experiment A1 assesses feature importance using SHapley Additive exPlanations (SHAP) to evaluate each feature's contribution to PRISM predictions. Experiment A2 evaluates the same global retrieval algorithm within different strata: Arctic sectors, latitude bands, and sea ice concentration regimes. Experiment A uses the baseline dataset (n=1,066,808 pixels); see Figure 3 and Table 1.

**Table 1**: Experiment A: input features set (30 features)

| Sensor/Source | Feature ID | Description |
|---|---|---|
| **SAR (S1)** | s1_hh, s1_hv | HH co-polarised and HV cross-polarised backscatter |
| | s1_cr | Cross polarisation difference (s1_hh - s1_hv) |
| | s1_hv_norm | Normalised HV (s1_hv/s1_hh) |
| **Passive microwave (AMSR2)** | TB_18V, TB_18H | Brightness temperature at 18 GHz, vertical and horizontal pol |
| | TB_36V, TB_36H | Brightness temperature at 36 GHz, vertical and horizontal pol |
| | TB_89V, TB_89H | Brightness temperature at 89 GHz, vertical and horizontal pol |
| | PR_18, PR_36, PR_89 | Polarisation ratios at each frequency ((V − H) / (V + H) ) |
| | GR_36_18V, GR_89_36V, GR_89_18V | Gradient ratios between frequency pairs (V-pol) |
| | TB_89V_36V_diff, TB_89H_36H_diff, TB_36V_18V_diff | Channel differences |
| | icecon_frac | Sea ice concentration fraction |
| **SST (OISST)** | sst_C | Sea surface temperature |
| | sst_above_freeze | SST minus seawater freezing point |
| **Cross-sensor interactions** | S1_hh_x_icecon, s1_hv_x_icecon ,pr18_x_icecon | SAR HH, SAR HV and PR_18 × ice concentration |
| **Spatiotemporal** | doy_sin, doy_cos | Cyclical day-of-year encoding |
| | month | Calendar month |
| | row_norm | Normalised grid row (latitude proxy) |
| | col_norm | Normalised grid column (longitude proxy) |

***Experiment B - Sensor and feature contribution studies***

Experiment B is designed to quantify the independent contribution of each sensor and feature group to retrieval performance. To do this, the same model is retrained under eight input configurations (B1-B8; Table 2), each removing, isolating, or substituting specific feature groups relative to the full baseline.

**Table 2:** Experiment B - Sensor and features configurations

| Configuration | Features | Pixel | Purpose |
|---|---|---|---|
| **B1**<br>**All features** | 35<br>(30 base + 5 CryoSat) | 499,046 | All features from baseline PRISM retrieval algorithm/Experiment A including CryoSat-2 freeboard |
| **B2**<br>**Physical only** | 30 (Removed spatiotemporal) | 499,046 | Test sensors alone without location or data information |
| **B3**<br>**Spatiotemporal only** | 5 | 499,046 | Tests pure climatological location-season map with no sensor data |
| **B4**<br>**No SAR** | 28 (removed 4 SAR features and the interaction terms) | 499,046 | Mimics a passive-microwave-only retrieval: removes all SAR features including SAR×SIC interaction terms |
| **B5**<br>**No interactions** | 32 features (removed 3 interaction features) | 499,046 | Tests whether SAR × ice concentration interactions add skill beyond using each independently |
| **B6**<br>**No ice concentration** | 31 (removed icecon_frac + 3 interactions) | 499,046 | Tests the combined contribution of ICECON-derived features |
| **B7**<br>**No CryoSat,**<br>**CS pixels** | 30 (base features only) | 499,046 | Direct fair comparison against A to isolate the CryoSat-2 contribution |
| **B8**<br>**No CryoSat,**<br>**full dataset** | 30 (base features only) | 1,066,808 | Reference showing baseline performance on the complete dataset without any CS filtering |

Experiment B allows to isolate the incremental value of each sensor or sensor combination and to quantify whether the model is primarily learning from physical sensor signals or from spatiotemporal patterns in the data. All configurations B1-B7 were run on the CryoSat-available pixel dataset (CS-training dataset; 499,046 pixels), comprising cases where a valid 14-day CryoSat-2 composite exists, to ensure fair comparison between configurations with and without CryoSat, on identical pixel population. Configuration B8 uses the full baseline dataset (1,066,808 pixels) as an additional reference for performance without CryoSat-2 or pixel filtering based on CS availability.

***Experiment C - Spatial specialisation analysis***

Experiment C assesses spatial specialisation using the baseline dataset (n=1,066,808 pixels, 30 base features), consistent with Experiment A. Experiment C1 trains separate models for each stratum: Arctic sector, latitude band, and ice concentration regime, using only pixels from that stratum and evaluated under LOYO cross-validation within that stratum. Performance is then compared between the stratum-specific retrieval and the pan-Arctic PRISM retrieval algorithm to determine whether specialisation provides an advantage over a single global retrieval algorithm. Experiment C2 evaluates feature importance within each stratum to assess whether dominant predictors are consistent across regions or whether different regions rely on different signals.

# 5. Results

## 5.1. Experiment A1: Performance of baseline PRISM retrieval

Experiment A1 evaluates the baseline PRISM retrieval, trained on the 30 input features listed in Table 1, across eight melt seasons (2017-2025, excluding 2019) under Leave-One-Year-Out (LOYO) cross-validation. The cross-validation framework, which evaluates the retrieval algorithm on melt seasons entirely withheld from training, provides an independent assessment of the retrieval algorithm's ability to generalise to unseen conditions. The mean per-fold performance (where each fold corresponds to one held-out melt season) is RMSE=0.080 with a $R^2$=0.468, giving equal weight to each of the eight held-out years. These metrics quantify how closely the generated MPF dataset, PRISM-MPF, agrees with the OLCI-Istomina reference product across the pan-Arctic domain, and are used throughout this study to assess sensor contributions and regional skill. Computed over all 1,066,808 pixels simultaneously, the pooled out-of-fold metrics show better results (RMSE=0.077 and $R^2$=0.532), as they are dominated by the two largest years (2017 and 2018) and also incorporate between-year variance in mean MPF, which the per-fold mean does not. Nevertheless, throughout this study, the per-fold mean is used as the primary performance metric to ensure each melt season/year contributes equally to the assessment. Per-year $R^2$ ranged from 0.431 (2018) to 0.559 (2023), being stable across seven of the eight held-out years (Fig. 5). The 2024 year is a clear outlier: it is the smallest fold (n=25,932), attributable to the combined effect of reduced S1 Arctic observations following the loss of S1B in December 2021 and degradation in S1A acquisitions after April 2024 until the launch of S1C in December 2024. The exclusion of this year raises $R^2$ to 0.502 suggesting that performance across typical melt seasons is more stable than the full eight-fold mean suggests.

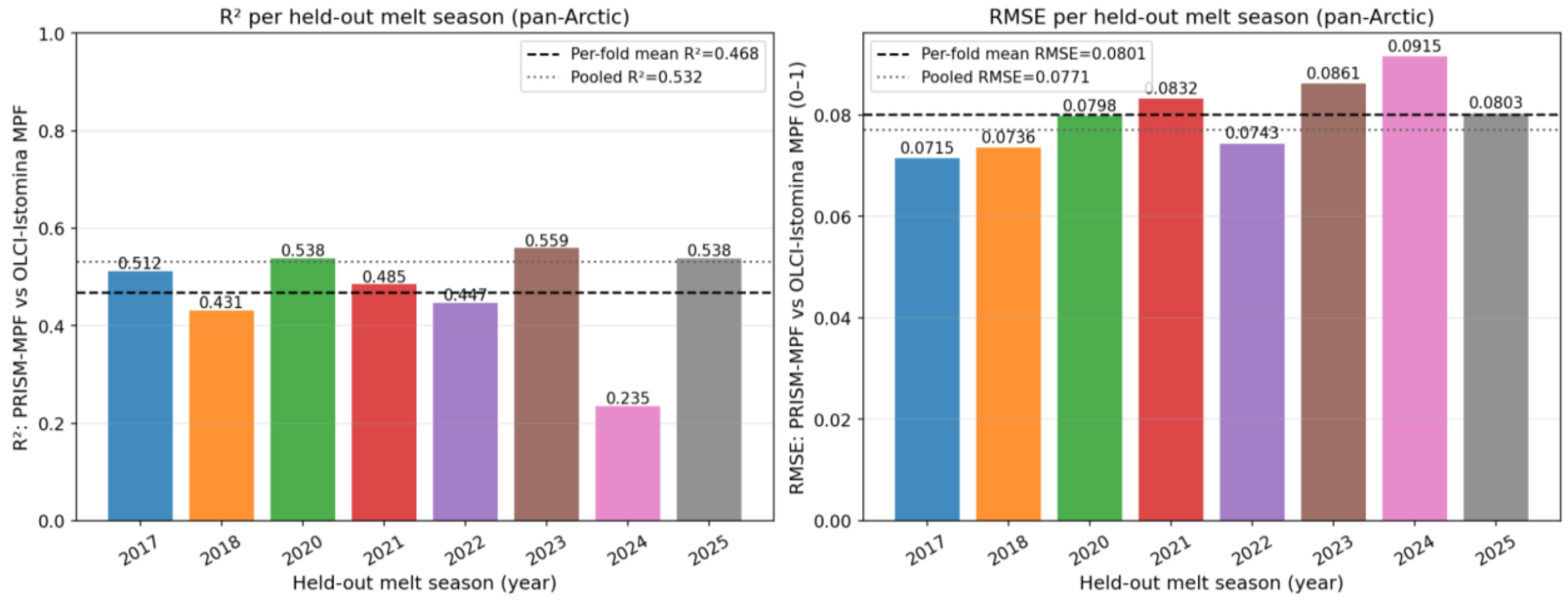

**Figure 5:** Year-by-year performance of the baseline PRISM-MPF retrieval under LOYO cross-validation

The PRISM-MPF (predicted MPF) versus OLCI-Istomina (target MPF) scatter plot (Fig.6, left) shows most of the predictions concentrated along the 1:1 line for MPF values in the range 0.0-0.4, which reflects the high density of results in this range (which makes up 95% of the dataset). RMSE remains low and relatively stable below MPF=0.3 range and increases markedly above MPF=0.4 (Fig.6, right). The deterioration in RMSE matches the lower pixel count, suggesting that the loss of retrieval skill is primarily attributable to insufficient training samples for heavily ponded conditions (MPF > 0.4) rather than a fundamental sensor limitation.

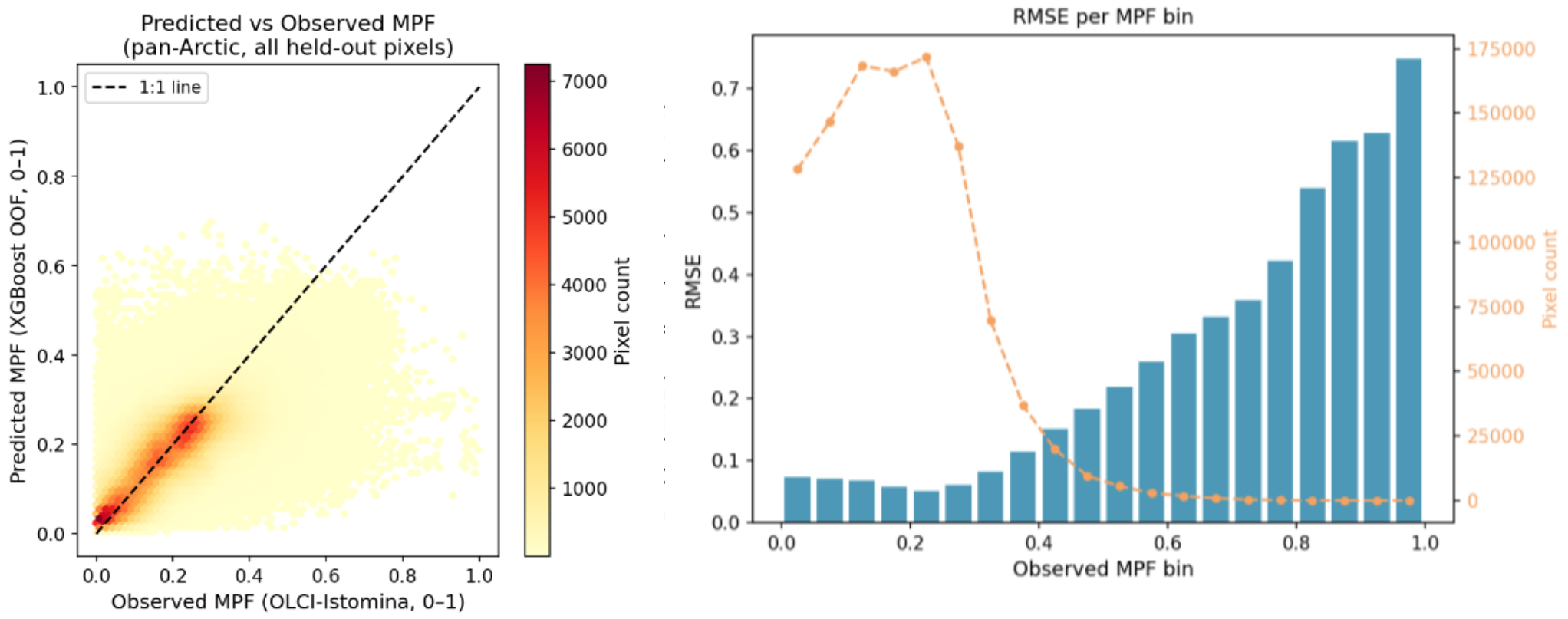

**Figure 6:** Baseline retrieval algorithm skill and error distribution by MPF range. Left: density-weighted scatter plot of out-of-fold predicted versus observed melt pond fraction (MPF) across all 1,066,808 held-out pixels under Leave-One-Year-Out cross-validation. Right: RMSE per 0.05-width MPF bin alongside pixel count per bin.

Looking at performance at the monthly level, the PRISM-MPF tracks the seasonal pattern across all eight melt seasons, but with residual underestimations of peak July values in anomalously high-MPF years, most evident in 2020 and 2023 (Fig. 7). Monthly RMSE is lowest in June and highest in August, reflecting a gradual loss of the retrieval skill as pond drainage and spatial heterogeneity intensify later in the season. Two monthly outliers with negative $R^2$ are visible in Fig. 7: July 2020 ($R^2 \approx -0.74$) and August 2017 ($R^2 \approx -0.39$). July 2020 is the most pronounced, attributable to the Arctic experiencing a significant anomaly in summer melting, leading to the smallest ever recorded July sea ice coverage (23.1% below the 1981-2010 average extent) in the 42-year satellite record (Fetterer et al., 2025). During July 2020, this anomaly was driven by persistent low pressure over the central Arctic, leading to exceptionally warm conditions and rapid sea ice melt in the Kara, Laptev and East Siberian Seas (Liang et al., 2022). Negative $R^2$ occur when the squared prediction errors exceed the variance of the OLCI-Istomina MPF values in that period (see Eq.10), and those two outlier months happened for different reasons. July 2020 was among the months with the highest mean MPF of the record, while at the same time was one of the months with the narrowest spread of MPF values across the pixels (i.e. standard deviation ≈0.07 derived from the monthly $R^2$ and RMSE), meaning that in this month, most pixels had similarly high pond cover (around 0.31 ± 0.07), whereas in the other months of same year they varied more among them (e.g. June's standard deviation ($\sigma \approx 0.10$)). This low variance made $R^2$ highly sensitive to errors, which were also moderately larger than in the surrounding months (RMSE=0.094, compared with 0.075 and 0.079), with PRISM-MPF systematically underestimating MPF). Conversely, August 2017 had a typical spread of MPF values (standard deviation ≈0.09). Its negative $R^2$ on the other hand reflects a systematic overestimation of MPF together with one of the highest monthly RMSE values in the record (0.104). In both months, about 30% of the squared error arises from a systematic offset (the squared mean bias; Eq. 12), with the remaining pixel-level differences being similar in magnitude to the within-month variance of the target MPF. In these anomalous months, the variation of MPF among individual pixels and days within the month was more difficult to reproduce than in typical melt seasons.

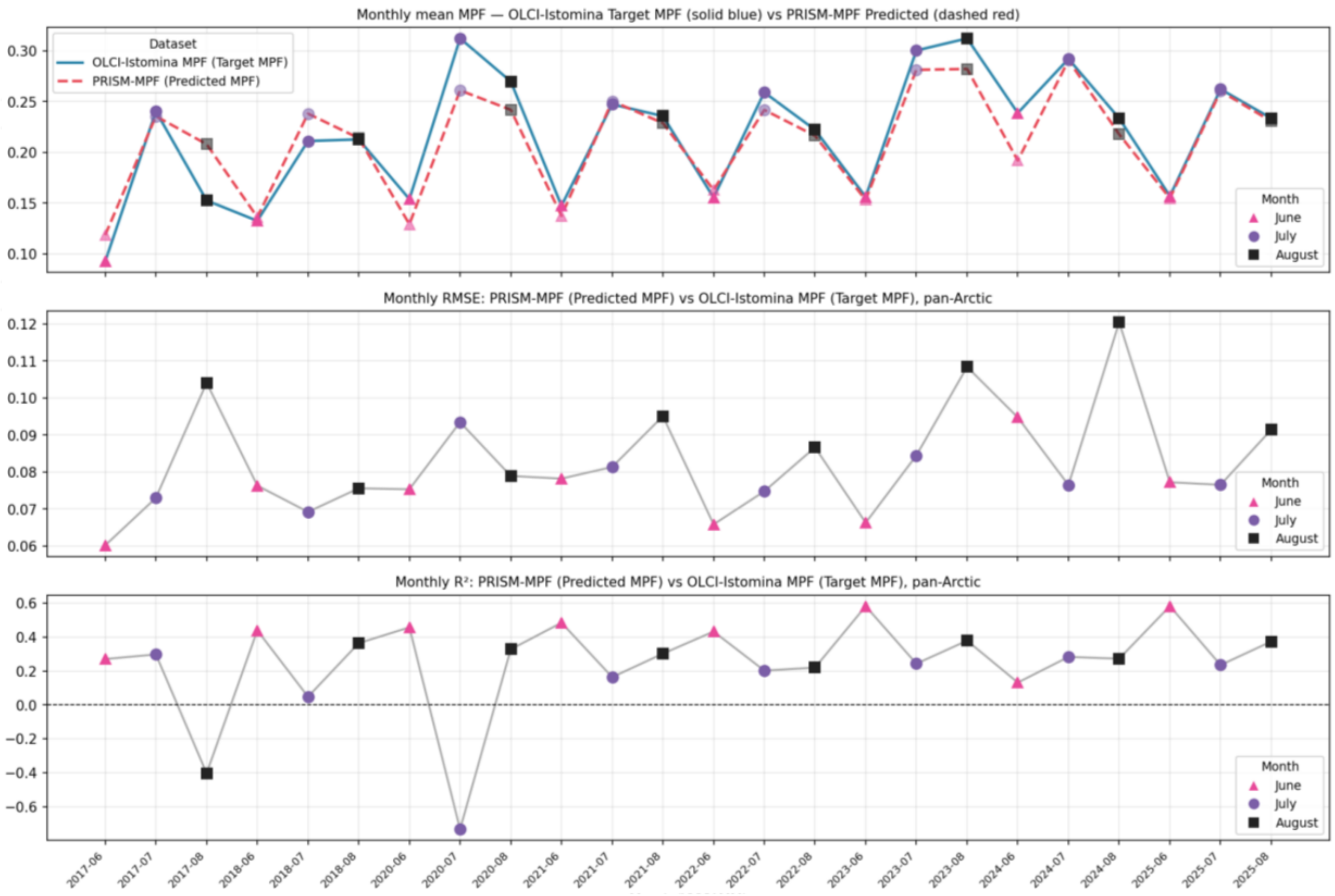

**Figure 7:** Monthly performance of MPF prediction across all melt seasons. **(a)** Monthly mean MPF from the OLCI-Istomina reference (target MPF, solid blue line) and from PRISM-MPF (predicted MPF, dashed red line). **(b)** Monthly RMSE of PRISM-MPF relative to the OLCI-Istomina reference, pan-Arctic. **(c)** Monthly $R^2$ of PRISM-MPF relative to the OLCI-Istomina reference, pan-Arctic; the dashed horizontal line marks $R^2 = 0$. Marker shapes indicate the month: June (triangles), July (circles) and August (squares).

Figure 8 illustrates a spatial example of the PRISM-MPF predictions for one of the best-performing dates, identified under the combined selection criterion (lowest RMSE with $R^2 \geq 0.40$, minimum 5,000 pixels). This selected date, 12 June 2018 (with an RMSE=0.059 and $R^2$=0.664) showcases a typical June pattern, with a mean low MPF=0.114 (from the reference OLCI-Istomina product), representative of the early pond expansion phase. The right panel indicates the residual (difference between PRISM-MPF and OLCI-Istomina target MPF), with blue and red indicating under and over predictions, respectively. To note that 21 dates with lower absolute RMSE were excluded by the $R^2 \geq 0.40$ quality gate, the majority from early June 2017.

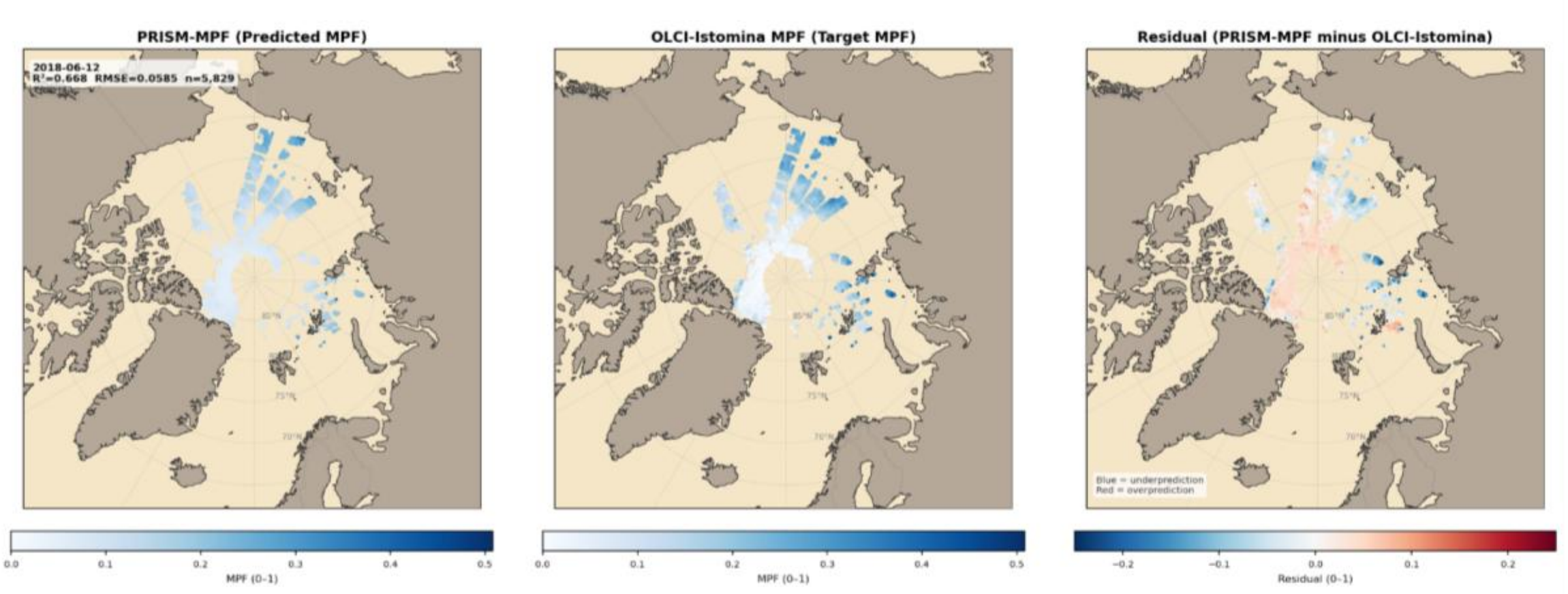

**Figure 8:** Pan-Arctic  mid-June (2018-06-12) showing: predicted MPF, observed MPF, and residual (pred minus obs)

The mean SHAP values analysis (left plot of Fig. 9) shows that the three top-ranked features are spatiotemporal: day-of-year (*doy_sin*=0.299) and pixel column and row (*col_norm*=0.127 and *row_norm*=0.112). These are followed by sea surface temperature (*sst_C*=0.093) and SAR HV polarisation (*s1_hv*=0.072), with calendar month (*month*=0.065) ranked sixth. The next highest-ranked physical predictors are the 18 GHz V-polarisation brightness temperature (*TB_18V*=0.056) and the cross-polarisation difference (*s1_cr*=0.055). Below this, the remaining 22 predictors, which span passive-microwave $T_B$ differences, gradient ratios, and SAR/ice-concentration predictors, have substantially smaller values, running from 0.003 to 0.038. The $T_B$ channels in particular span 0.003-0.014, most appearing in the lower half of the ranking, confirming that $T_B$ channels at 12.5 km resolution have a modest individual contribution to PRISM predictions. This is consistent with known constraints on passive microwave MPF retrieval: individual $T_B$ channels are natively coarser than typical target grid resolutions (Tanaka et al., 2016), and higher-frequency channel combinations have been shown to carry a narrower dynamic range and be less sensitive to MPF variation individually (Tanaka and Scharien, 2022).

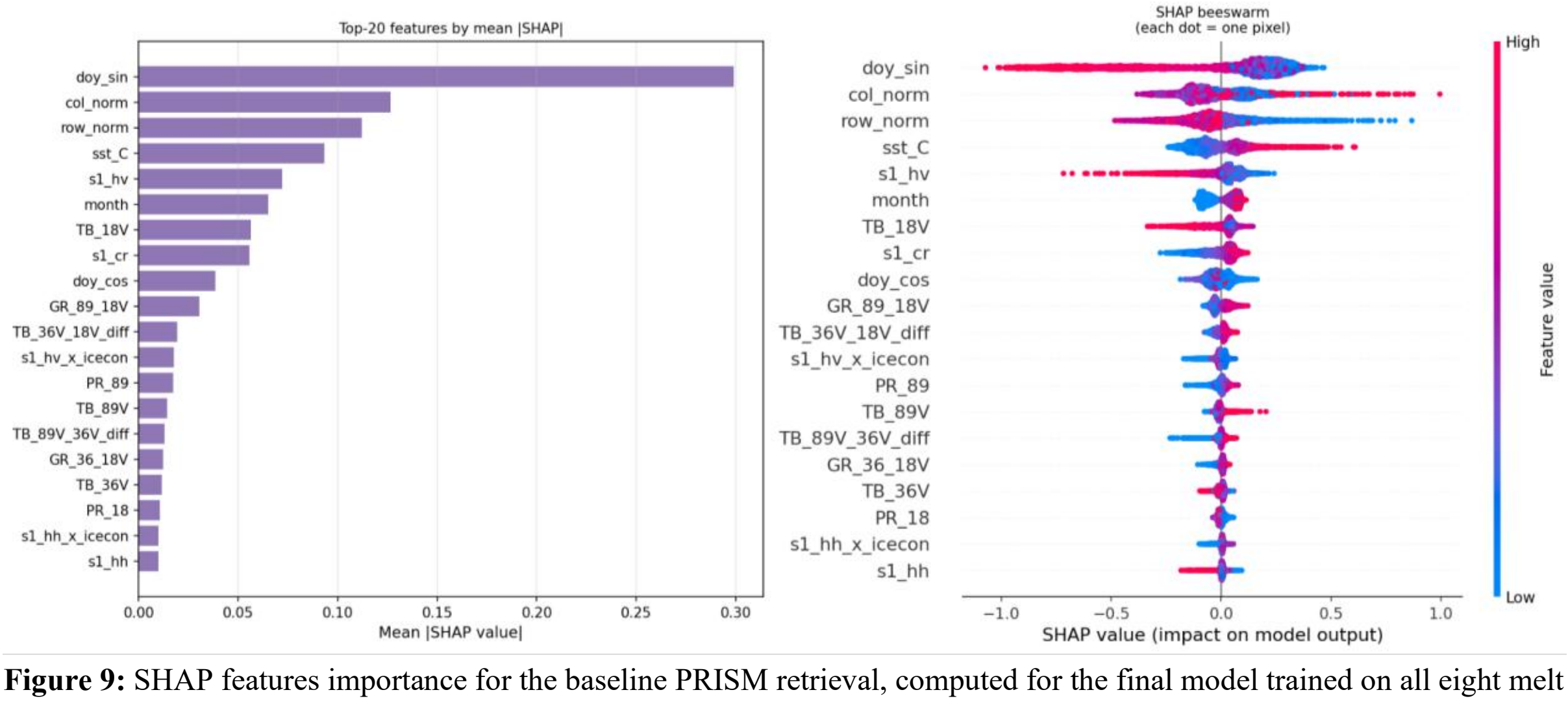

**Figure 9:** SHAP features importance for the baseline PRISM retrieval, computed for the final model trained on all eight melt seasons and based on a 5,000-pixel subsample. Left: top 20 features ranked by mean |SHAP value|. Right: SHAP beeswarm plot, where each dot represents one pixel's SHAP value for a given feature, coloured by that feature's value (red = high, blue

= low); position along the x-axis shows the feature's contribution to pushing the predicted MPF up (positive) or down (negative) relative to the model's average output

The SHAP Beeswarm plot (right chart of Fig. 9) explains the contribution of each feature to the PRISM predictions by pushing the predicted values up (positive SHAP values) or down (negative SHAP values) relative to the mean predicted MPF. Each dot represents one pixel's SHAP value for a given feature. For instance, sea surface temperature (*sst_C* in the plot) shows a near-monotonic relationship: high SST values (red dots) are associated with positive SHAP values, pushing predictions towards higher MPF, while lower SST (blue dots) push predicted values down. However, since OISST SST under sea ice is derived from ice concentration (Section 3.3), this relationship partly reflects ice conditions rather than temperature alone. Moreover, passive microwave sea ice concentration retrievals may interpret melt ponds as open water during the melt season (Banzon et al., 2020), so the proxy SST may itself partly encode pond presence. A similar but inverse pattern is visible for SAR HV polarisation (*s1_hv*): high cross-polarised backscatter values (red dots) push towards lower MPF, while low backscatter (blue dots) is associated with positive SHAP values, meaning higher predicted MPF. This is consistent with SAR HV cross-polarisation backscatter behaviour, with smooth surfaces (such as melt ponds under calm wind conditions) producing weaker backscatter due to specular reflection, while dry rough surfaces (such as bare ice) tend to produce stronger surface scattering, resulting in higher HV backscatter. So lower *s1_hv* values coincide with higher MPF, which is consistent with expected radar signature of melt pond pixels. While SAR backscatter-related features (*s1_hv, s1_cr*) show a wide spread of SHAP values with clear positive and negative swings across pixels, the 18 GHz V-polarisation brightness temperature (*TB_18v*) shows a visibly narrower, more compact distribution. This contrast suggests the two channel types are capturing different scales of variability, possibly linked to each sensor's ability to resolve features at different spatial resolutions.

### 5.2. Regional stratification of baseline retrieval performance (Experiment A2)

Building on the baseline PRISM retrieval algorithm (Experiment A; Fig. 5), we next assessed whether its performance holds uniformly across space and time. Experiment A2 stratifies the out-of-fold PRISM-MPF retrievals by three physical strata: Arctic sectors (described in Fig. 2), latitude bands (85-90°N; 80-85°N; 75-80°N; 70-75°N; and 60-70°N), and sea ice concentration regime (Very High >70%; High 40-70%; Moderate 15-40%; and Low <15%). This allows us to understand whether the retrieval skill varies systematically across different regions or sea ice concentration regimes.

The spatial stratification by the Arctic sector (Fig. 10, top) reveals a considerable heterogeneity in retrieval skill. The CAA, Beaufort Sea, and Central Arctic show the highest skill ($R^2$=0.544, 0.530, and 0.526, respectively), close to the global pooled $R^2$=0.532, with only the CAA exceeding it. The poorest performance is observed in the Laptev Sea and East Siberian Sea ($R^2$=0.220 and $R^2$=0.280, respectively), with both also showing the highest absolute errors (RMSE > 0.10), whilst the three best-performing sectors achieve RMSE values below 0.09. Stratification by latitude band (Fig. 10, bottom left) shows a monotonic decline in skill towards lower latitudes: $R^2$=0.530 at 85-90°N, falling to $R^2$=0.317 at 70-75°N, with $R^2$=0.019 at 60-70°N. Correspondingly, RMSE increases from 0.070 at the highest latitudes towards to lowest latitudes, though the 60-70°N band should be interpreted with caution given its very small sample size. Stratification by sea ice concentration regime (Fig. 10, bottom right) shows the same pattern in a different dimension: $R^2$=0.530 in the Very High ice concentration regime (>70%, n=999,911, comprising approximately 94% of all pixels), declining to $R^2$=0.384 in the Low ice concentration regime (<15%), closest to the ice edge.

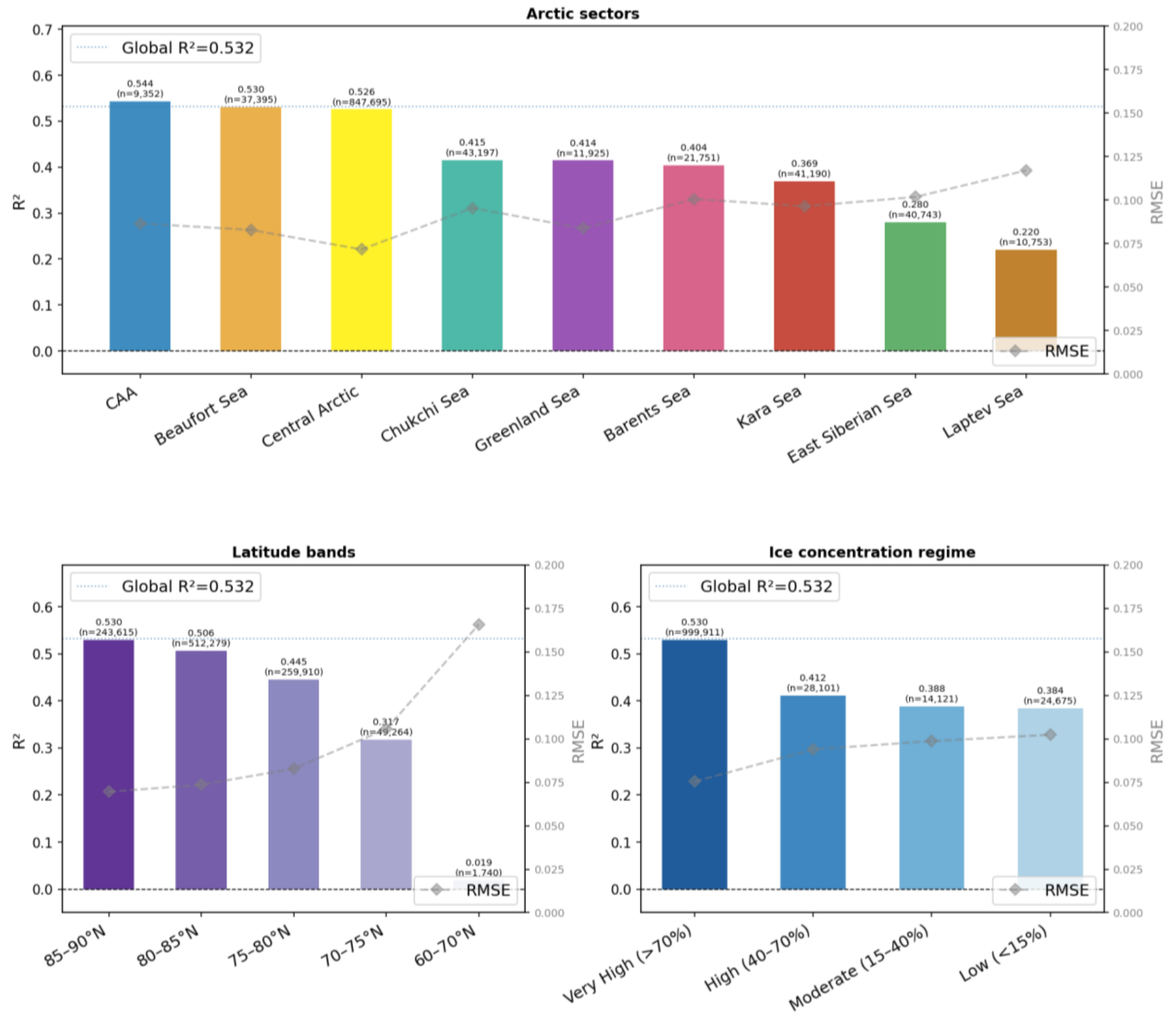

**Figure 10:** Spatial stratification of retrieval performance. The pan-Arctic baseline PRISM retrieval (leave-one-year-out cross-validation; LOYO CV) was evaluated within subsets of its pooled out-of-fold (OOF) predictions, grouped by (top) Arctic sector, (bottom left) latitude band, and (bottom right) sea ice concentration regime. Bars show $R^2$ (left axis, coloured) and diamonds connected by a dashed line show RMSE (right axis, grey) for each stratum; sample size (n) is annotated above each bar. The dotted horizontal line marks the global pooled-OOF $R^2$ (0.532) for comparison.

### 5.3. Experiment B: Sensor and feature contribution studies

Following Experiment A, which established a global baseline retrieval algorithm, whose skill varies systematically across space, Experiment B was designed to quantify the contribution of each sensor and input group to that skill, and whether the full sensor suite used is necessary to achieve it (Fig. 4). The same XGBoost model architecture is retrained under a series of eight different input configurations (B1-B8; Table 2), each removing, isolating, or substituting specific feature groups relative to the full baseline; B1 being the reference configuration with all 35 features trained on the CryoSat-available pixel subset (n=499,046). The configurations comprise: physical features alone (B2, $R^2$=0.375) and spatiotemporal features alone (B3, $R^2$=0.261), to decompose the main SHAP feature classes; all features (including CryoSat-2 data) with SAR removed (B4, $R^2$=0.380) to mimic a passive-microwave-only retrieval; interaction terms removed (B5, $R^2$=0.463) to test whether SAR × ice-concentration interactions add skill beyond their individual components; and ice concentration-derived features removed (B6, $R^2$=0.464).

All configurations B1-B7 are trained and evaluated on the identical CryoSat-available (CS) pixel subset (n=499,046 pixels, comprising those where a valid 14-day CryoSat-2 composite exists), ensuring configurations with and without CryoSat-2 are compared on the same pixel population. B7 removes CryoSat-2 features while keeping the same CS pixel subset as B1-B6, directly isolating CryoSat-2's contribution (B7, $R^2$=0.450).

Configuration B8 uses the same feature set as B7 (no CryoSat) but on the full baseline dataset (n=1,066,808), serving as a reference for performance without any CS-driven pixel filtering. The CryoSat-2 cost/benefit trade-off is evaluated through two complementary comparisons: B1 vs. B7 isolates CryoSat's contribution while holding pixel count constant, showing what CryoSat-2 alone adds; B7 vs. B8 reflects the practical trade-off between restricting to CryoSat-valid pixels (n≈500K) versus retaining the full dataset (n≈1M) without CryoSat-2. All eight configurations are evaluated under LOYO cross-validation, consistent with Experiment A. Two quantities are reported. The standalone skill of a feature group is the $R^2$ obtained when only that group is used (Fig. 11). Its removal contribution is the $\Delta R^2$ lost when the group is removed from the full configuration B1 (Table 3). They answer different questions and need not rank groups identically.

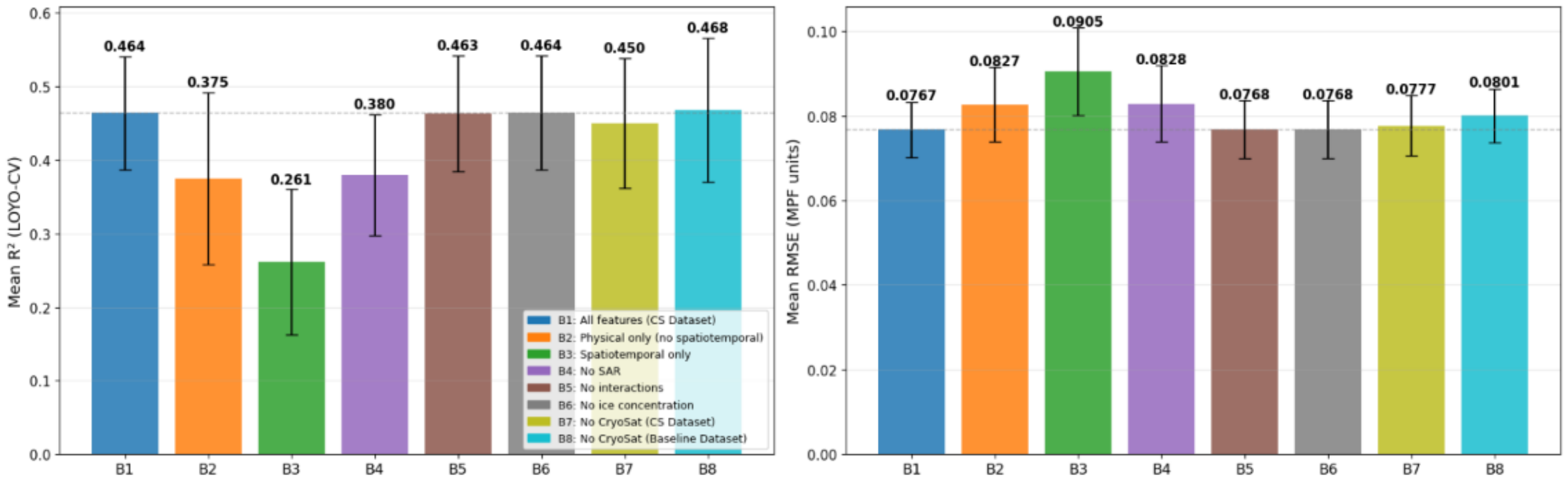

**Figure 11:** Experiment B (configurations B1-B8, Table 2): mean $R^2$ (left) and mean RMSE (right) across LOYO cross-validation folds. Dashed line marks B1 (all features) as reference. B1-B7 use the CS-training dataset (n=499,046, CryoSat-2-valid pixels only); B8 uses the same features as B7 (i.e. excluding CryoSat-2) but on the full baseline dataset (n=1,066,808). Error bars show ±1 standard deviation across the eight LOYO folds.

Considering standalone skill first (Fig. 11), physical sensors alone (B2, $R^2$=0.375) outperformed spatiotemporal features alone (B3, $R^2$=0.261) by $\Delta R^2$≈+0.114 (Table 3), indicating that physical sensor data carry predictive power beyond location and season alone. In terms of removal contributions (Table 3), removing the spatiotemporal features (B1−B2) lowered skill by $\Delta R^2$=+0.089, roughly a fifth of the full configuration's $R^2$ (B1, $R^2$=0.464), whereas removing all physical sensors (B1−B3) cost $\Delta R^2$=+0.203. Spatiotemporal and physical features are therefore complementary rather than redundant. Removing SAR alone (B1−B4) produced a loss comparable to that of the spatiotemporal group ($\Delta R^2$=+0.085), disproportionately large given that SAR accounts for only four of 35 features. This is the first pan-Arctic quantification of C-band SAR's independent value for MPF retrieval. By removing all SAR-derived features, including the SAR×SIC interaction terms, configuration B4 provided a clean passive-microwave-only baseline. HV polarisation (*s1_hv*) is the most important individual SAR feature, consistent with its sensitivity to surface and volume scattering changes associated with melt pond presence. In contrast, removing cross-sensor interaction terms (B5) and ice-concentration-derived features (B6) had negligible effect (both $|\Delta R^2|<0.002$), indicating that sea ice concentration (*icecon_frac*) and its interaction terms carry little independent predictive value once location, season, and SAR backscatter are already included.

The CryoSat-2 comparison requires two distinct tests (Table 3). Holding pixel count fixed, CryoSat-2 inclusion (B1 vs. B7) yields a modest gain ($\Delta R^2$=+0.014). However, restricting to CryoSat-valid pixels discards 53% of the dataset (499,046 vs. 1,066,808 pixels). The pixel-count gain alone (B7 vs. B8, both without CryoSat, $\Delta R^2$=+0.018) is larger than CryoSat's own contribution, and the full-dataset configuration without CryoSat-2 (B8, $R^2$=0.468) outperforms the CryoSat-inclusive configuration on the restricted pixel subset (B1, $R^2$=0.464) overall. Taken together, these results indicate CryoSat-2 carries genuine but modest predictive information, which is outweighed by the cost of the restricted number of collocated pixels that configuration B1 requires.

**Table 3.** Key pairwise comparisons from the Experiment B configurations

| Comparison | ΔR² | Interpretation |
|---|---|---|
| B1 − B2 (all features − no spatiotemporal) | +0.089 | Spatiotemporal features' total contribution to retrieval skill |
| B1 – B3 (all features −spatiotemporal only) | +0.203 | Physical sensor's total contribution to retrieval skill |
| B2 − B3 (physical only − spatiotemporal only) | +0.114 | Physical sensors add skill beyond pure climatology |
| B1 − B4 (all features − no SAR) | +0.085 | SAR's independent contribution; comparable to spatiotemporal |
| B4 − B3 (no SAR − spatiotemporal only) | +0.119 | Passive microwave, SST and CryoSat-2 add skill beyond climatology |
| B1 − B5 (all features − no interactions) | +0.001 | Cross-sensor interaction terms add negligible independent skill |
| B1 − B6 (all features − no ice concentration) | +0.0002 | Ice-concentration features carry no independent skill beyond existing predictors |
| B1 − B7 (all features − no CryoSat, same n) | +0.014 | CryoSat-2's information gain, holding pixel count fixed |
| B8 − B7 (baseline dataset − CS-training dataset, both no CryoSat) | +0.018 | Gain from the larger dataset alone, independent of CryoSat |
| B8 − B1 (baseline dataset, no CryoSat − CS-training dataset, with CryoSat) | +0.004 | Net effect: more data outweighs CryoSat's inclusion |

## 5.4. Experiment C1: Spatial specialisation analysis - spatial partitioning

Whilst Experiment A establishes a global baseline retrieval algorithm and Experiment B shows how its skill is driven by different predictors (with spatiotemporal and SAR features contributing most among individual sources), Experiment C1 addresses whether training a model specifically on each region's own data performs better than the global retrieval algorithm. The same spatial strata defined in Experiment A2 (i.e., by Arctic sectors, latitude bands, and ice concentration regimes), each with its own LOYO cross-validation, are used to test whether specialising on a single spatial context outperforms the global retrieval algorithm. Experiment C2 employs a per-stratum feature importance heatmap to assess whether the predictors driving PRISM retrieval skill are consistent across regions or vary spatially.

For comparability, both local and global $R^2$ values in Experiment C1 are computed as pooled out-of-fold metrics on identical pixel populations within each stratum. The global $R^2$ for each stratum is derived from the pooled out-of-fold predictions of the global retrieval algorithm evaluated on that stratum's pixels, the same predictions used in Experiment A2 stratification diagnostics, rather than the per-fold mean reported as the headline metric in Experiment A1 ($R^2$=0.468). The two metrics differ because pooled OOF weights pixels equally whilst the per-fold mean weights years equally; global $R^2$ values within individual strata therefore differ slightly from the pan-Arctic per-fold mean.

Across all 17 strata which were evaluated (9 Arctic sectors, 4 latitude bands and 4 ice concentration regimes; Fig. 12), local specialisation did not improve performance relative to the global retrieval algorithm in any stratum. The CAA came closest to parity ($\Delta R^2$=−0.005; local $R^2$=0.539 versus global $R^2$=0.544), the smallest loss among the Arctic sectors despite being the smallest sector by pixel count (n=9,352). Degradation was generally larger in more marginal strata: among the Arctic sectors, the largest drops occurred in the Laptev Sea ($\Delta R^2$=−0.126, local $R^2$=0.094) and Barents Sea ($\Delta R^2$=−0.070), both amongst the smaller sector populations (n=10,753 and n=21,751, respectively). However, sample size and degradation are not strictly linked, as the CAA result already shows: the 85-90°N band showed the largest degradation among the latitude bands ($\Delta R^2$=−0.076) despite a comparatively large sample (n=243,615). On the other hand, the largest stratum overall, very high ice concentration (>70%, n=999,911), showed the smallest degradation of any stratum ($\Delta R^2 = -0.002$), consistent with the general trend that larger, more data-rich strata are least affected by the loss of cross-stratum training information. These results indicate that training on the full pan-Arctic dataset outweighs any benefit from spatial specialisation across all strata.

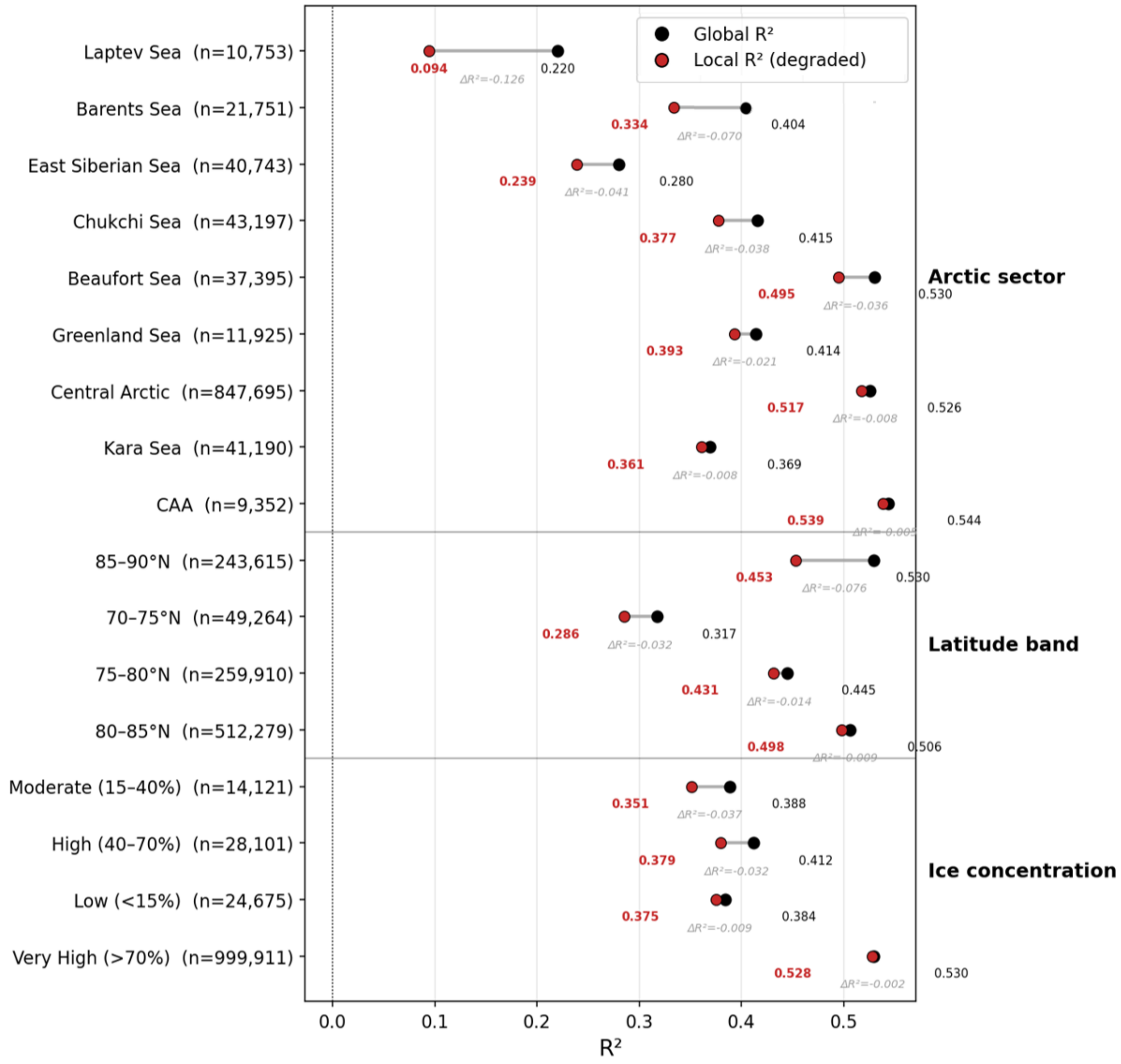

**Figure 12.** Local vs Global $R^2$ across all strata

## 5.5. Experiment C2: Per-stratum feature importance

Also for this experiment, we computed feature importance, to understand if the predictors driving retrieval skill are consistent across strata, or instead whether different regions would show a specific signature. In contrast to Experiment A, where SHAP was computed for the global retrieval algorithm, here the XGBoost model is trained and evaluated exclusively within each stratum, such that feature importance reflects the local microwave-MPF relationship within that spatial context rather than the pan-Arctic signal. As a result, feature importance varies across strata. Amongst the sensor/physical features, SAR HV-polarisation backscatter (*s1_hv*) was the highest-ranked physical sensor predictor in the East Siberian Sea and Laptev Sea (second and third overall, respectively), whilst it was less important in the Central Arctic, Greenland Sea and CAA, where other features dominated. In the Beaufort Sea, the 89/18 GHz gradient ratio (*GR_89_18V*) emerged as the dominant physical predictor (second overall, after month), despite ranking only fifth amongst physical predictors at pan-Arctic level (Experiment A; Fig. 9). Because the 89 GHz channel is far more affected by scattering and liquid water than the 18 GHz channel, this ratio is sensitive to melt state, snow wetness and microstructure, surface temperature, ice type and the presence of open water within the satellite footprint and therefore carries information on the physical condition of the ice surface during summer melt (Kern et al., 2016; Tanaka and Scharien, 2022). The 18 GHz vertical-polarisation brightness temperature (*TB_18V*) was the highest-ranked physical sensor feature in the CAA, and the SAR cross-polarisation difference (*s1_cr*, the difference between HH and HV backscatter) in the Central Arctic; in both cases spatiotemporal features dominated overall. Sea surface temperature (*sst_C*), which ranked fourth at pan-Arctic level (Fig. 9), was elevated across most sectors, most strongly in the Barents Sea, but not in the Central Arctic, CAA and Beaufort Sea.

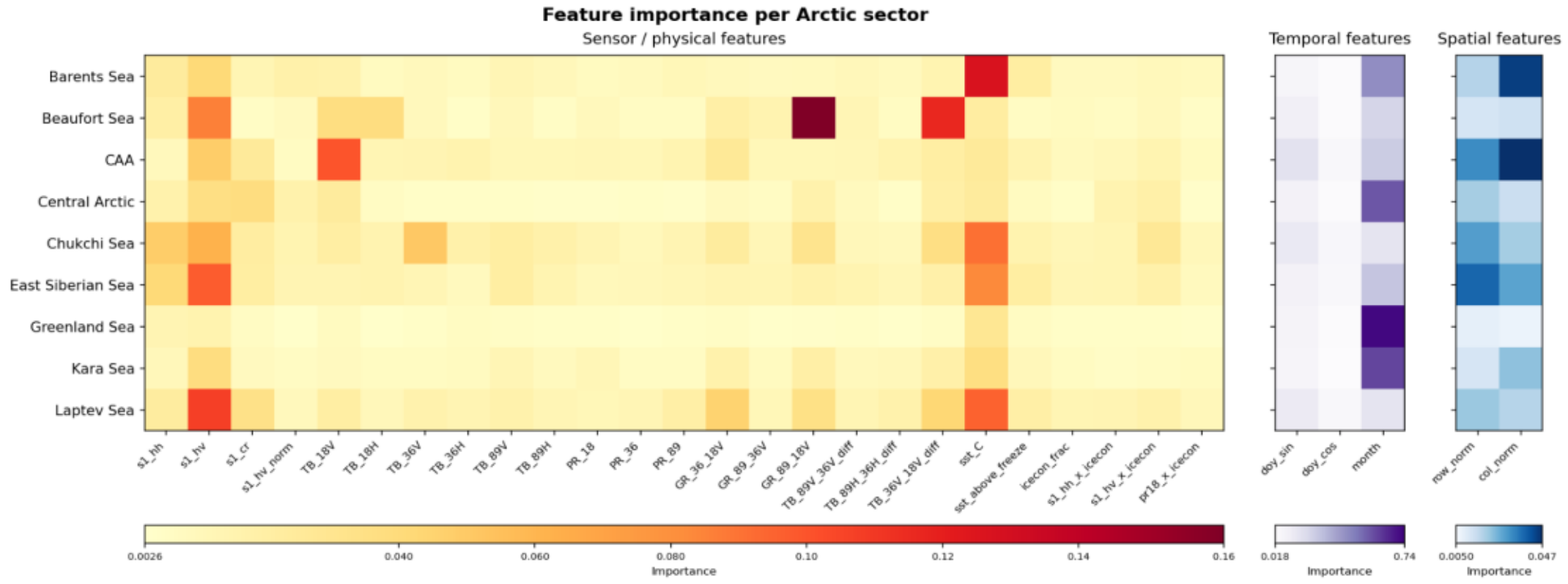

**Figure 13.** Feature importance per Arctic sectors

By latitude band (Fig. 14, upper panel), *s1_hv* importance was most pronounced in the 75-80°N band and remained notable in the 70-75°N and 80-85°N bands, while the 60-70°N band (n=1,740) showed a distinct pattern dominated by cyclical day-of-year encoding (*doy_sin*, *doy_cos*) rather than month, with the 36-18 GHz V-polarisation channel difference (*TB_36V_18V_diff*) and *sst_C* as the leading physical sensor features. By ice concentration regime (Fig. 14, lower panel), *s1_hv* importance was substantially higher in the Low, Moderate and High concentration regimes than in the Very High (>70%) regime, where sensor/physical feature importance was generally low, with *sst_C* as the highest-ranked physical feature.

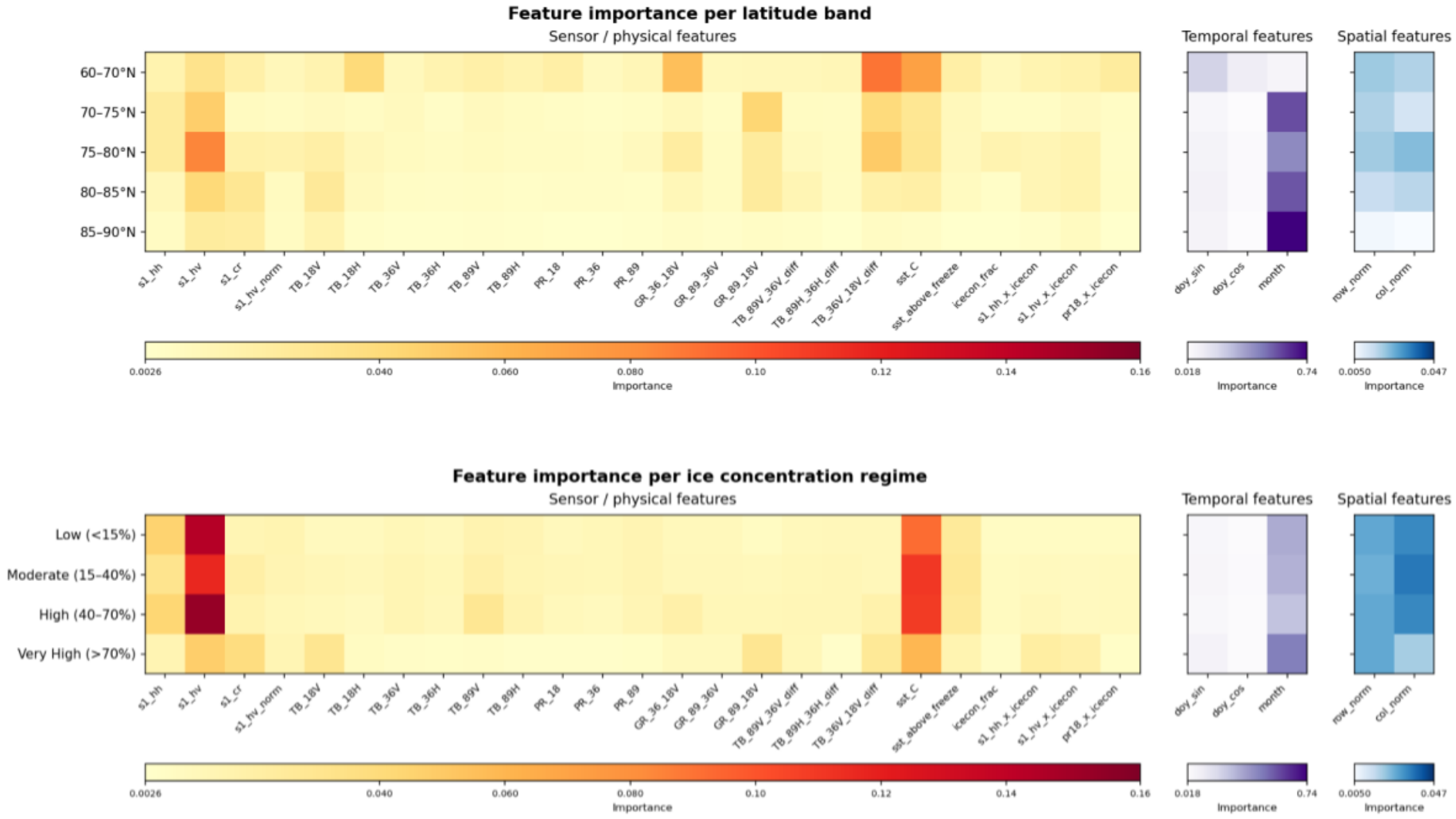

**Figure 14.** Feature importance per latitude bands (upper) and ice concentrations (lower)

Temporal feature importance (month) was high across all sectors, ice concentration regimes and latitude bands except 60-70°N (Fig. 13 and 14), and was highest in the Greenland Sea among sectors, in the 85-90°N band, and in the Very High ice concentration regime. Among ice concentration regimes, the regime with the strongest reliance on month (Very High) also showed the highest $R^2$, but this association does not hold across sectors or latitude bands: the Greenland Sea, for example, showed the strongest reliance on month despite intermediate skill. Spatial feature importance includes the normalised grid column and row position, which are used as longitude and latitude proxies (*col_norm*, *row_norm*). These spatial features have comparatively low importance overall, but show localised peaks: *col_norm* importance was distinctly elevated in the CAA and Barents Sea relative to the other sectors, and in the Low, Moderate and High ice concentration regimes relative to Very High; across latitude bands, *row_norm* was elevated between 60 and 80°N and *col_norm* peaked in the 75-80°N band, with both lowest at 85-90°N. Overall, predictor reliance is only partially consistent across strata: month ranks amongst the most important predictors in every stratum except the small 60-70°N band, while *sst_C* is consistently prominent in ice concentration regimes and several Arctic sectors but less prominent across latitude bands. SAR HV backscatter shows the greatest spatial inconsistency of any single predictor, dominant in some sectors and negligible in others.

## 6. Discussion

The PRISM-MPF dataset, a pan-Arctic, multi-year, all-weather melt pond fraction record derived from multi-source microwave and auxiliary satellite observations, demonstrates that machine learning-based sensor fusion can produce physically consistent MPF estimates across eight melt seasons, with skill broadly comparable to previous regionally constrained microwave retrievals. The baseline PRISM retrieval achieved stable multi-year performance under leave-one-year-out (LOYO) cross-validation (Experiment A). Feature importance analysis (Experiment A1) and systematic sensor ablation (Experiment B) established the contributions of spatiotemporal predictors, SAR, ice concentration, and CryoSat-2 features. Retrieval skill varied substantially across Arctic sectors, latitudes, and ice concentration regimes (Experiment A2), but regional specialisation did not improve performance (Experiment C1).

### 6.1. Retrieval skill and inter-annual stability

The conservative LOYO cross-validation approach, allows to confirm that PRISM achieves predictive skill for unseen years/melt seasons, ensuring no within-year memorisation is taking place. RMSE was stable across eight melt seasons (2017-2025, excluding 2019), indicating a consistent prediction accuracy. $R^2$ showed more variability, which reflects its sensitivity to the variance of MPF from the reference/target dataset, the OLCI-Istomina within each held-out year, rather than considerable differences in absolute retrieval error, an important interpretive distinction when evaluating the retrieval performance and resulting PRISM-MPF predictions.

The inter-annual stability of PRISM skill, consistent across seven of eight melt seasons, confirms that the baseline PRISM retrieval algorithm has learned a robust pan-Arctic representation of the microwave-MPF relationship that generalises across diverse melt conditions. Two cases are good examples to showcase why $R^2$ and RMSE divergence do not necessarily imply a genuine breakdown in accuracy. First example is the 2024 held-out year: its RMSE (0.092) is not considerably higher than the multi-year mean (0.08), while its $R^2$ (0.235) is a considerable outlier. This year had a considerably smaller pixel count (n=25,932), so its sample is less representative of pan-Arctic conditions than other seasons. A second example is July 2020 (Fig. 7), which experienced the lowest Arctic sea ice extent for that month in the satellite record (Liang et al., 2022). In that month, the spread of OLCI-Istomina MPF was among the narrowest in the record, whilst mean MPF was among the highest (Section 5.1). The negative $R^2$ therefore reflects both a compressed target variance, which amplifies the effect of any error on $R^2$, and a moderate loss of accuracy, with RMSE rising to 0.094 and PRISM-MPF underestimating the unusually high pond cover. These two cases show that isolated low or negative $R^2$ values should not, on their own, be read as a breakdown in retrieval accuracy, while July 2020 also provides a useful test case for anomalous melt conditions beyond the training distribution, a point revisited in Section 6.7.

### 6.2. Interpretation of retrieval skill against a reference-product ceiling

The interpretation of the results of this study's performance metrics, requires two distinct comparisons: against the reference product's own uncertainty, and against prior retrieval studies. The target reference variable of this study, the OLCI-Istomina MPD2 product, is an optically-derived retrieved product with its own associated uncertainty. The authors reported $R^2$=0.52 and RMSE=0.14 against in situ observations (Istomina et al., 2015a), while validation against aerial imagery over high-concentration pack ice yielded a lower RMSE (0.065); while the more recent MPD2 algorithm reports an uncertainty of approximately 7.8% (Niehaus et al., 2024). By comparison, PRISM achieved a per-fold RMSE=0.080 ($R^2$=0.468) under LOYO validation (and $R^2$=0.532, RMSE=0.077 when computed on pooled out-of-fold prediction (Section 5.1). This study's RMSE should therefore be interpreted as a measure of agreement with the OLCI-Istomina reference, i.e., how well PRISM-MPF reproduces an existing optical product that itself carries non-trivial uncertainty, rather than as an estimate of absolute MPF accuracy. Importantly, unlike optical retrievals, which require daytime and cloud-free conditions, the approach developed here operates independently of illumination and cloud cover, offering continuous all-weather estimates at pan-Arctic scale across different years. Like any retrieval trained with optical observations, PRISM assumes that the microwave-MPF relationship learned under clear skies holds under cloud cover, an assumption that future in situ or airborne campaigns timed to overcast conditions could directly test. The fact that the current performance achieved sits within the range of reference-product uncertainty, despite fundamental differences, indicates that this approach delivers usable retrievals.

### 6.3. Positioning against prior relatable retrieval studies

Taking into account the reference-product ceiling established above, comparisons with other studies presented below should be interpreted as a relative positioning with the broader MPF retrieval literature rather than an absolute ranking. A direct comparison cannot be straightforward, and this is, to the best of our knowledge, the first pan-Arctic microwave-based MPF retrieval study, following a LOYO CV protocol. Prior microwave-based studies have made crucial contributions within spatially confined domains (single regions, specific ice type regimes or campaign-based measurements. PRISM builds directly on these regional insights while expanding the spatial and temporal capabilities. Its results (RMSE=0.080), is comparable in order of magnitude to previous SAR-based and passive-microwave studies, while extending to the full a wider, pan-Arctic and interannual domain.

Among SAR-based retrievals, Scharien et al. (2014) achieved an RMSE=0.05-0.07 using a dual-polarisation (VV/HH) SAR model over landfast FYI near CAA during a single melt season. Ramjan et al. (2018) achieved, with their best model, RMSE=0.09 ($R^2$=0.62) using winter C-band RADARSAT-2 linear polarimetric data to predict early-summer MPF over landfast FYI in Resolute Passage in 2012, testing a range of incidence angles with far-range performing best. Scharien et al. (2017) achieved RMSE=0.09 predicting spring MPF in the CAA from winter SAR backscatter. Each of these studies operated under temporally and spatially confined conditions, as opposed to pan-Arctic conditions.

Among passive-microwave retrievals, Tanaka et al. (2016) developed an AMSR-E passive microwave MPF algorithm and reported an RMSE of 8.9% against ship-based MPF observations from the HOTRAX2005 and JOIS2011 cruises in the Beaufort Sea and Canadian Arctic Archipelago, with differences against MODIS optical MPF generally under 5% in most regions and periods; the algorithm was restricted to areas of high sea ice concentration (>95%). Tanaka & Scharien (2022) evaluated higher-resolution AMSR-E/AMSR2 channel combinations, as an alternative to the original coarse 6.9 GHz-based retrieval, achieving standard deviations of ~3.8% relative to the original product over the Canadian Arctic Archipelago (2006-2018), though assessed against product-variant agreement rather than independent ground-truth validation.

As expected, optical-based studies achieve lower errors overall, given their spectral discrimination capabilities and since they are not subject to the same fundamental physical limitations as microwave sensors due to coarse spatial resolution and ambiguity in discriminating water surfaces (ponds, leads, open ocean) in both SAR and passive microwave signatures. Peng et al. (2022) achieved RMSE=0.05 ($R^2$=0.76) using MODIS data from 2000-2020 with a genetic-algorithm-optimized back-propagation neural network (GA-BPNN). Ding et al. (2020)

achieved RMSE=0.083 (comparable to this study's per-fold RMSE), using MODIS data from 2000-2019 with an ensemble deep neural network under clear-sky conditions. Lee et al. (2020) achieved RMSE=0.18 using MODIS band differences. PRISM performance falls within the range spanned by several optical studies, whilst operating under all-weather conditions, the greatest limitation of optical retrievals. This represents a meaningful advance for microwave-based methods: it demonstrates that fusing multiple microwave sources with machine learning can produce all-weather MPF estimates of comparable quality to established optical methods, even as optical retrievals retain an accuracy advantage in absolute terms under clear-sky conditions. Moreover, Experiment B addresses a question that no prior microwave-based MPF study has attempted: the quantification of each sensor group's independent contribution to pan-Arctic retrieval skill. The significance of these results is discussed in the following section.

### 6.4. Quantification of independent sensor contributions at pan-Arctic scale

Experiment B goes beyond prior microwave-based MPF studies by decomposing, for the first time, the relative value of each input source, physical sensors versus spatiotemporal context, in driving pan-Arctic PRISM skill. The results reveal that spatiotemporal features and C-band SAR make comparable removal contributions, while physical sensors as a group carry more standalone skill than spatiotemporal features, with direct implications for the design of future all-weather MPF retrieval systems.

Among individual input sources, spatiotemporal features and C-band SAR make the largest and comparable removal contributions ($\Delta R^2$ = +0.089 and +0.085; B1–B2 and B1–B4, Table 3), while removing all physical sensors together costs $\Delta R^2$ = +0.203 (B1–B3). However, when used alone, spatiotemporal features yield substantially lower skill (B3: $R^2$ = 0.261, RMSE = 0.091) than physical sensors alone (B2: $R^2$ = 0.375, RMSE = 0.083), with a difference in $R^2$ of approximately 0.114. This contrast indicates that, although spatiotemporal features substantially improve the full model, physical sensor data provide considerable standalone predictive power beyond climatological location-season information. The two groups of input sources are thus complementary rather than redundant. Their complementarity is also visible in the year-to-year stability of the retrieval, not just its mean skill: the spread of $R^2$ across LOYO folds is noticeably wider when either feature family is isolated (B2, B3) than in the full configuration (B1; error bars in Fig. 11). This suggests spatiotemporal and physical features do not simply add together in mean skill, in fact they appear to stabilise each other's performance across interannual variability, both groups contributing to consistent generalisation to unseen years.

The C-band SAR independent contribution is worth highlighting. Removing SAR (B1-B4; Table 3) produced a drop broadly comparable in magnitude to the spatiotemporal contribution itself despite SAR accounting for only four of 35 features. HV polarisation is the most important SAR feature, probably due to its sensitivity to surface and volume scattering changes, providing stronger contextual ice information related to melt ponds than passive microwave. Although SAR has been used for melt onset and melt pond detection (e.g. Han et al., 2016; Mahmud et al., 2016; Maknun et al., 2024), this is the first quantification of C-band SAR's incremental value beyond passive microwave in a pan-Arctic, multi-year framework. This finding directly motivates the incorporation of L-band SAR (ALOS-2/PALSAR-2, NISAR) in future PRISM-MPF iterations: L-band's longer wavelength penetrates more deeply into wet snow (Casey et al., 2016) and is less sensitive to wind-driven surface roughening, providing complementary information precisely in the calm-wind, early-season conditions where C-band spectral discrimination is weakest. The negligible contributions of cross-sensor interaction terms ($\Delta R^2$=+0.001) and ice concentration-derived features ($\Delta R^2$=+0.0002) are a useful simplification result: these features can be omitted from future PRISM implementations without meaningful skill loss, reducing model complexity without sacrificing performance. The redundancy of the ice concentration features is likely partly explained by OISST SST itself, which under sea ice is derived from ice concentration (Section 3.3), so this information remains available to PRISM through *sst_C* even when the ice concentration features are removed.

With respect to the CryoSat-2, the better performance of the full-dataset configuration without CryoSat-2 (B8, $R^2$=0.468) relative to the CryoSat-inclusive restricted pixel subset configuration (B1, $R^2$=0.464) is not a dismissal of radar altimetry as a potential PRISM-MPF input, but a consequence of current swath geometry and an informative finding about the nature of CryoSat-2's MPF signal. CryoSat-2's radar altimeter is known to be directly affected by melt ponds during the summer melt season: pond surfaces produce specular (mirror-like) radar returns that bias the retrieved surface elevation and complicate the ice/lead classification the sensor's standard processing relies on (Kwok et al., 2018; Dawson and Landy, 2023). Because of this bias, conventional CryoSat-2 processing has historically been restricted to winter months, with reliable summer-melt-season retrievals only recently becoming feasible through specialised processing (Landy et al., 2022). This suggests CryoSat-2 freeboard is not simply a noisy or indirect proxy for MPF, but is directly physically responsive to melt pond presence, via the same specular-reflection mechanism that also complicates the retrieval of sea ice thickness. However, this responsiveness does not appear to translate into a strong, independently exploitable predictive signal for MPF once SAR backscatter, passive microwave, and spatiotemporal features are already included. This is plausible because the specular-return signature captured by CryoSat-2 overlaps substantially with what SAR backscatter already captures more directly and at finer spatial scales. CryoSat-2's narrow altimeter swath imposes a 53% pixel cost that the modest information gain ($\Delta R^2$=+0.014) cannot justify at present. This is a constraint of the sensor's orbital geometry and sub-satellite sampling rather than of the retrieval approach and is therefore not addressable through better feature engineering or model tuning. ESA's dual satellite HydroGNSS Scout reflectometry mission and the planned CRISTAL Sentinel Expansion mission with its dual-frequency Ku/Ka-band altimeter, may add complementary new all-weather products in future. Alongside Sentinel-1 these are natural candidates for incorporation into future PRISM-MPF versions, potentially changing the cost-benefit balance identified here substantially.

### 6.5. How the retrieval algorithm resolves MPF: spatiotemporal context and sensor complementarity

The convergence of SHAP rankings, ablation results and stratification diagnostics supports a consistent interpretation of how PRISM resolves MPF across the pan-Arctic domain. At feature level, spatiotemporal variables dominate the SHAP ranking (Fig. 9); at group level, they make a removal contribution ($\Delta R^2$=+0.089) comparable to that of C-band SAR ($\Delta R^2$=+0.085) but are weaker standalone than physical sensors (B3 vs B2). PRISM thus draws on two complementary sources: a seasonal and geographic baseline, and locally observed surface conditions. C-band SAR provides the strongest independent microwave contribution (Section 6.4), while CryoSat-2 adds modest skill on matched pixels. Although AMSR2 brightness temperatures generally rank lower in SHAP (except TB_18V; Fig. 9), AMSR2, SST and CryoSat-2 jointly improve performance beyond spatiotemporal features alone (B4 vs B3, $\Delta R^2$=+0.119).

These differences between sensors reflect their spatial resolution: SAR operates at finer native resolution and responds to surface scattering at the pond scale, whereas passive microwave brightness temperatures integrate over a much coarser footprint and are not directly sensitive to sub-pixel pond structure. Whether this pan-Arctic structure holds at the regional scale, or whether different regions rely on different sensors, is examined in the following section.

### 6.6. Pan-Arctic generalisation versus regional specialisation

Experiment C1 tests whether the spatial heterogeneity in the baseline PRISM-MPF retrieval skill, which varies systematically across Arctic sectors, latitude bands, and ice concentration regimes, could be reduced by training dedicated regional models. The results of such spatial specialisation analysis have shown that regional specialisation negatively affects MPF prediction performance in every case: a single pan-Arctic retrieval algorithm (Experiment A) outperforms region-specific retrievals across all 17 Arctic sectors, latitude bands and ice concentration regimes (Results, Fig. 12). This is connected to the importance of spatiotemporal features in Experiments A1 and B: splitting the dataset by sector, or by latitudes removes the spatial diversity encoded by the normalised latitude and longitude positions proxies (*row_norm, col_norm,* Table 1) upon which the retrieval algorithm relies the most. When the size of the training data is drastically reduced, a double penalty is imposed which region-specific retrievals cannot overcome.

The regions with the poorest performance of the global retrieval algorithm are the Laptev Sea and the East Siberian Sea, and local training degraded both further. For the Laptev Sea, this is plausibly linked to its small sample (n=10,753). The East Siberian Sea (n=40,743), however, is comparable in size to the Kara, Chukchi and Beaufort Seas, which all perform better, suggesting that data volume alone does not explain its weak skill and that regional conditions not captured by the current predictors may also contribute. Conversely, the CAA came closest to parity under sector-specific training ($\Delta R^2 = -0.005$, the smallest loss amongst the sectors), despite being the smallest sector by pixel count (n=9,352). The CAA is geographically distinct from all the others, being also a reason for being repeatedly subject of dedicated SAR-MPF retrievals (Yackel & Barber, 2000; Scharien et al., 2017; Ramjan et al., 2018; Howell et al., 2020). This result suggests that spatial specialisation may become worth considering in regions where landfast ice is common, provided that longer records supply sufficient local training data. Such fast ice can support relatively stable melt-pond conditions even as broader sea-ice concentration declines over the season.

At pan-Arctic scale the two feature groups contribute complementarily (Section 6.5), but the relative importance of individual physical sensors is not fixed across regions. SAR backscatter, month, and (to a lesser extent) spatial position each dominate differently across regions, rather than one sensor consistently playing this role everywhere. This could be linked to the physical interpretation developed earlier: SAR's discriminating power depends on wind-driven surface roughening and ice/water backscatter contrast, both of which vary systematically with regional ice conditions (e.g. *s1_hv* importance is highest in the Low, Moderate and High ice concentration regimes, and in the East Siberian and Laptev Seas, where it is the leading physical predictor, and comparatively low where ice cover is uniform and consolidated (Central Arctic, Very High ice concentration)). In the latter more homogeneous regions, the retrieval algorithm instead falls back more heavily on the seasonal (month) signal, since a distinctive backscatter-based signature is less available. This suggests the retrieval algorithm does not apply a single fixed sensor hierarchy across the Arctic but rather adapts which physical signal it draws on according to the discriminating information locally available. This regional signal diversity also could point to where future sensor additions would potentially have the greatest impact: L-band SAR, with deeper penetration and reduced wind-roughening impact, could specifically improve retrieval performance.

### 6.7. PRISM: first generation foundation capabilities and future extensions

The current PRISM-MPF framework represents a first-generation retrieval and a foundation, from which refinements in design and sensor extensions could result in improvements in future PRISM-MPF iterations. Several aspects of this study should be acknowledged given the span of inherited limitations coming from sensor ambiguity, training data volume and model architecture, alongside potential future work that could address them. Some of the characteristics described below are not fundamental barriers to the approach, but specific, addressable aspects of the current implementation, each allowing future iterations of PRISM-MPF framework to address them.

SAR backscatter sensitivity to melt ponds, is not purely a function of the ice/water dielectric contrast, but modulated as well by wind-driven surface roughening (Scharien et al., 2007). Melt ponds under calm conditions produce weak, largely specular backscatter that is difficult to distinguish from surrounding ice, whereas wind-roughened pond surfaces scatter more strongly and become substantially more separable from ice in HH and HV backscatter; wind speed has been shown to explain the majority of variance in this pond-backscatter relationship (Yackel & Barber, 2000). If the SAR SHAP contributions identified in Section 5 partly reflect this wind-roughening effect rather than pond presence alone, this would suggest that incorporating surface wind speed (from scatterometers, numerical weather predictions, or SAR-derived wind speed) as an explicit predictor/input variable, could add independent predictive value by improving discrimination between calm-water specular returns from ice-covered surfaces; while simultaneously addressing the atmospheric forcing sensitivity described below.

Atmospherically forced anomalies represent another potential iteration to consider. The July 2020 event (Section 5.1), characterised by record-low sea ice extent and a PRISM-MPF underestimation of the unusually high pond cover, is a valuable test case for how PRISM performs when atmospheric forcing drives melt conditions beyond the historical training distribution. Adding an explicit atmospheric predictor such as surface wind speed or a circulation index, could jointly address the wind-roughening ambiguity discussed above while improving the ability of the PRISM retrieval algorithm to adapt to anomalous melt regimes. As the observational records will extend beyond 2025, the training volume will broaden naturally.

Training data volume and class imbalance (which regard underrepresented conditions) is another strategy element. RMSE increased sharply when MPF>0.4 (Fig. 6), a data scarcity constraint rather than a sensor limitation. Similarly, the weakest-performing spatial strata (Laptev Sea, East Siberian Sea) are characterised by smaller pixel populations rather than fundamentally different microwave-MPF relationships. These gaps are addressable through targeted oversampling of high-MPF conditions, an MPF-weighted loss function, or continued accumulation of training years as the observational record extends.

Regarding model architecture and spatial resolution, the co-location process imposed a 12.5 km grid, which resulted in discarding the benefit of SAR's much finer native resolution. Moreover, each pixel is treated independently, preventing the model from exploiting spatial context, (i.e. from neighboring pixels), which could be physically meaningful, since melt ponds are not randomly distributed but form following surface structures which are potentially visible at fine SAR resolution. Retaining SAR at near or equal to its native resolution, or a convolutional approach capable of learning from spatial neighbourhood context, could exploit this structure whilst simultaneously increasing effective training sample size by operating at higher resolution. This represents arguably the single most impactful architectural improvement available within the current sensor constellation.

Beyond addressing these near-term improvements, further extensions could test the broader generalisability of the PRISM-MPF framework. Applying it to alternative optical reference products, such as MODIS-LEE (Lee et al., 2020; Lee & Stroeve, 2021) or MODIS-SADIKNI (Sadikni & Kern, 2026), would clarify whether the sensor contributions identified here are specific to OLCI-Istomina or holds across independently derived datasets. Further improvements in sensors, model, training data and reference datasets would strengthen PRISM-MPF's capability for all-weather Arctic monitoring.

## 7. Conclusions

This study presents PRISM-MPF, the first all-weather, pan-Arctic, multi-year MPF dataset, generated by the PRISM retrieval framework from an XGBoost model trained on multi-source microwave and auxiliary satellite observations and evaluated under a rigorous cross-validation protocol across eight melt seasons (2017-2025, excluding 2019). The baseline PRISM-MPF retrieval algorithm achieves a per-season mean $R^2$=0.468 and RMSE=0.080 across 1,066,808 co-located pixels. This performance sits within the uncertainty range of the OLCI-Istomina reference product and is comparable in order of magnitude to previous regionally specific microwave-based studies.

Among individual input sources, spatiotemporal features ($\Delta R^2$=+0.089) and C-band SAR backscatter ($\Delta R^2$=+0.085) make the largest and broadly comparable removal contributions to PRISM skill, the latter representing the first pan-Arctic quantification of SAR's independent contribution to MPF retrieval. Physical sensors alone nevertheless outperform spatiotemporal features alone ($R^2$=0.375 vs 0.261), showing that the two groups are complementary rather than redundant. These results are consistent with both SHAP feature importance analysis (Experiment A1) and systematic sensor ablation (Experiment B). CryoSat-2 radar freeboard carries genuine but modest predictive information ($\Delta R^2$=+0.014); however, its sparse altimeter swath coverage restricts the training dataset to 499,046 pixels, discarding 53% of the full baseline, a pixel cost that outweighs the information gain. Ice concentration features add negligible independent skill. Among the microwave inputs, C-band SAR makes the largest independent contribution, while passive microwave and altimetry add complementary, regionally varying information.

PRISM skill varies substantially across Arctic sectors, latitude bands, and ice concentration regimes (Experiment A2), but this spatial heterogeneity is not addressable through regional specialisation: the pan-Arctic PRISM-MPF retrieval algorithm outperforms locally trained alternatives in every stratum (Experiment C1), confirming that the diversity of pan-Arctic training outweighs any local adaptation benefit.

PRISM-MPF is a first-generation all-weather microwave MPF dataset, cross-validated against the OLCI-Istomina optical reference product, whose all-weather, pan-Arctic retrieval skill is comparable to previous regionally constrained microwave studies. The framework is built to grow: directly extensible as new microwave sensors become available, with L-band SAR and wider-swath altimetry representing promising near-term additions, and with retrieval skill in anomalous melt years improving naturally as the training record extends beyond 2025. With successive improvements in sensor coverage, model architecture, training data volume and reference dataset quality, the PRISM framework provides a reproducible, extensible pathway towards continuous, all-weather Arctic MPF monitoring, a capability essential for understanding and predicting the Arctic's accelerating role in the global climate system.

## 8. Appendix A

Table A1: XGBoost hyperparameters

| Parameter | Value | Role |
|---|---|---|
| n_estimators | 800 | Number of boosting rounds |
| learning_rate | 0.03 | Shrinkage per tree (regularisation) |
| max_depth | 6 | Maximum tree depth |
| subsample | 0.8 | Row subsampling fraction per tree |
| colsample_bytree | 0.8 | Feature subsampling fraction per tree |
| min_child_weight | 10 | Minimum samples per leaf node |
| reg_alpha | 0.1 | L1 regularisation coefficient |
| reg_lambda | 1.0 | L2 regularisation coefficient |
| Random seed | 42 | Fixed for reproducibility |

## 9. CRediT authorship contribution statement

**Sara Aparício:** Writing (original draft, review & editing), Validation, Software, Methodology, Investigation, Formal analysis, Data curation, Visualization, Conceptualisation. **Simon Driscoll:** Writing (review & editing), Supervision, Methodology, Conceptualisation. **Mark R. Drinkwater:** Writing (review & editing), Supervision, Methodology, Conceptualisation. **Daniela Flocco:** Writing (review & editing), Supervision, Methodology, Conceptualisation.

## 10. Declaration of competing interest

The authors declare that they have no known competing financial interests or personal relationships that could have appeared to influence the work reported in this paper.

## 11. Declaration of AI-assisted technologies in the manuscript preparation process

The concept, analyses, results and interpretations were produced by the authors. During the preparation of this work, the authors used Claude mainly to identify typographical errors and polish the manuscript text and to verify that in-text citations matched the reference list. After using this tool, the authors reviewed and edited the content as needed and take full responsibility for the content of the published article.

## 12. Data availability

The input datasets used in this study are publicly available: Sentinel-1 GRD (Google Earth Engine collection COPERNICUS/S1_GRD), AMSR-E/AMSR2 AU_SI12 (Meier et al., 2018), NOAA OISST v2.1 (Huang et al., 2021), Cryo-TEMPO Summer Sea Ice (ESA, 2024) and OLCI-Istomina MPF (Istomina et al., 2025). The PRISM-MPF v1.0 dataset will be made publicly available in a repository with a permanent DOI upon acceptance for publication.

## 13. Acknowledgements

We are also grateful to the Portuguese Foundation for Science and Technology (Fundação para a Ciência e a Tecnologia - FCT) for supporting CENSE - Center for Environmental and Sustainability Research (UID/04085/2025 doi.org/10.54499/UID/04085/2025; UID/PRR/04085/2025 doi.org/10.54499/UID/PRR/04085/2025), and CHANGE - Institute for Global Change and Sustainability (LA/P/0121/2020 -doi.org/10.54499/LA/P/0121/2020).